\documentclass{article}

\usepackage[main, final]{neurips_2026}
 
\usepackage[utf8]{inputenc}
\usepackage[T1]{fontenc}
\usepackage{url}
\usepackage{booktabs}
\usepackage{amsfonts}
\usepackage{nicefrac}
\usepackage{xcolor}
\usepackage{amsmath}
\usepackage{amssymb}
\usepackage{mathtools}
\usepackage{amsthm}
\usepackage{graphicx}
\usepackage{bbm}
\usepackage{enumitem}
\usepackage{mathrsfs}
\usepackage{placeins}
\usepackage{makecell}

\definecolor{linkblue}{HTML}{1A5FB4}
\usepackage[
    colorlinks=true,
    linkcolor=linkblue,
    citecolor=linkblue,
    urlcolor=linkblue,
]{hyperref}

\newcommand{\numagents}{n}
\newcommand{\graphspace}{\mathfrak{G}}
\newcommand{\statespace}{\mathcal{S}}
\newcommand{\actspace}{\mathcal{A}}
\newcommand{\obsspace}{\mathcal{O}}
\newcommand{\transfunc}{\mathcal{T}}
\newcommand{\obsfunc}{\mathcal{U}}
\newcommand{\rewardfunc}{\mathcal{R}}
\newcommand{\horizon}{H}
\newcommand{\discount}{\gamma}

\newcommand{\state}{s}
\newcommand{\statet}{\state_t}
\newcommand{\statett}{\state_{t+1}}

\newcommand{\obs}{o}
\newcommand{\obst}{\obs_t}
\newcommand{\obstt}{\obs_{t+1}}

\newcommand{\act}{a}
\newcommand{\actt}{\act_t}

\newcommand{\reward}{r}

\newcommand{\rewardtt}{\reward_{t+1}}

\newcommand{\policy}{\pi}
\newcommand{\actobs}{\tau}
\newcommand{\actobst}{\actobs_t}

\newcommand{\envsym}{\phi}
\newcommand{\envsymset}{\Phi}

\newcommand{\graph}{\mathcal{G}}
\newcommand{\grapht}{{\mathcal{G}_t}}
\newcommand{\agentset}{N}
\newcommand{\edgeset}{E}

\newcommand{\gtruth}{\Psi}
\newcommand{\std}{\sigma}

\newcommand{\weight}{\alpha}
\newcommand{\accuracy}{T}
\newcommand{\conformity}{C}

\newcommand{\nullact}{\emptyset}

\newcommand{\identity}{\mathrm{Id}}

\newcommand{\interact}{\mathrm{I}}
\newcommand{\notinteract}{\mathrm{I'}}
\newcommand{\PrIt}{\Pr(\interact_t)}
\newcommand{\PrIo}{\Pr(\interact_0)}

\usepackage{acro}

\DeclareAcronym{marl}{
  short = MARL,
  long = multi-agent reinforcement learning,
}
\DeclareAcronym{rl}{
  short = RL,
  long = reinforcement learning,
}
\DeclareAcronym{zsc}{
  short = ZSC,
  long = zero-shot coordination,
}
\DeclareAcronym{egt}{
  short = EGT,
  long = evolutionary game theory,
}
\DeclareAcronym{lstm}{
  short = LSTM,
  long = long short-term memory,
}
\DeclareAcronym{decpomdp}{
  short = Dec-POMDP,
  long = decentralised partially observable Markov decision process,
}
\DeclareAcronym{nposg}{
  short = N-POSG,
  long = networked partially observable stochastic game,
}
\DeclareAcronym{aoh}{
  short = AOH,
  long = action-observation history,
}
\DeclareAcronym{sp}{
  short = SP,
  long = self-play,
}
\DeclareAcronym{op}{
  short = OP,
  long = other-play,
}
\DeclareAcronym{xp}{
  short = XP,
  long = cross-play,
}
\DeclareAcronym{abm}{
  short = ABM,
  long = agent-based model,
}
\DeclareAcronym{gru}{
  short = GRU,
  long = gated recurrent unit,
}
\DeclareAcronym{ppo}{
  short = PPO,
  long = proximal policy optimisation,
}
\DeclareAcronym{ippo}{
  short = IPPO,
  long = independent proximal policy optimisation,
}
\DeclareAcronym{iql}{
  short = IQL,
  long = independent Q-learning,
}
\DeclareAcronym{er}{
  short = ER,
  long = Erd\H{o}s--R\'{e}nyi,
}
\DeclareAcronym{ba}{
  short = BA,
  long = Barab\'{a}si--Albert,
}
\DeclareAcronym{dba}{
  short = DBA,
  long = directed Barab\'{a}si--Albert,
}
\DeclareAcronym{ws}{
  short = WS,
  long = Watts--Strogatz,
}
\DeclareAcronym{sbm}{
  short = SBM,
  long = stochastic block model,
}
\DeclareAcronym{awi}{
  short = AWI,
  long = attention weighted importance,
}
\DeclareAcronym{mse}{
  short = MSE,
  long = mean squared error,
}
\DeclareAcronym{ode}{
  short = ODE,
  long = ordinary differential equation,
}
\DeclareAcronym{bfs}{
  short = BFS,
  long = breadth-first search,
}
\DeclareAcronym{ne}{
  short = NE,
  long = Nash equilibrium,
}
\DeclareAcronym{psd}{
  short = PSD,
  long = private signal disagreement,
}
\DeclareAcronym{pi}{
  short = PI,
  long = perceived importance,
}
\DeclareAcronym{cce}{
  short = CCE,
  long = coarse correlated equilibrium,
}
\DeclareAcronym{llm}{
  short = LLM,
  long = large language model,
}

\usepackage[capitalize,noabbrev]{cleveref}

\title{Modelling Opinion Dynamics \\ at Scale with Deep MARL}

\author{%
  Lukas Seier\thanks{Correspondence to \texttt{lukas.seier@eng.ox.ac.uk}} \\
  BOLD \\
  University of Oxford \\
  \And
  Brandon Kaplowitz \\
  Witt Lab \\
  University of Oxford \\
  \And
  Sebastian Towers \\
  BOLD \\
  University of Oxford \\
  \AND
  Richard Bailey \\
  OUCE \\
  University of Oxford \\
  \And
  Jakob Foerster \\
  BOLD \\
  University of Oxford \\
}

\begin{document}

\maketitle

\begin{abstract}
Modelling opinion dynamics typically relies on hand-crafted local interaction rules to study emergent macroscopic phenomena such as consensus and polarisation. In contrast, multi-agent reinforcement learning (MARL) enables agents to learn such behaviours directly by optimising simple rewards. To explore the potential of MARL for opinion dynamics, we introduce a GPU-accelerated \textit{consensus and truth-finding game} that scales to populations of up to 1000 agents, comparable to many real-world social sub-networks. To prevent unrealistic conventions, we extend other-play to general-sum social interactions. We next validate our model on a subset of the Bluesky network by recovering agent importance structures from graph topology alone via a learned attention layer, finding that highly conforming populations most closely match human data. In large social media networks, such high levels of conformity significantly reduce collective accuracy and promote dishonest agents that lie to fit in. By contrast, small hunter-gatherer networks are less affected; here, conformity can even improve collective agreement. This suggests a mismatch between evolved human conformity heuristics and modern social media environments as a potential contributor to misinformation.
Our code is available at \url{https://github.com/flipbagels/OpiniMARL}.
\end{abstract}
\section{Introduction} \label{sec:introduction}
In the face of global challenges, such as climate change and health pandemics, identifying the forces that promote or inhibit collective agreement and truth finding is critical to sustaining functioning societies~\citep{ipcc2023synthesis,morens2020emerging}. Rapid growth of social media has popularised the study of opinion dynamics~\citep{starnini2025opinion}, exploring mechanisms responsible for large-scale phenomena, such as consensus formation, polarisation and pluralistic ignorance~\citep{baronchelli2018emergence,baumann2020modeling,centola2005emperor}.

\begin{figure}[t]
    \centering
    \includegraphics[width=\linewidth]{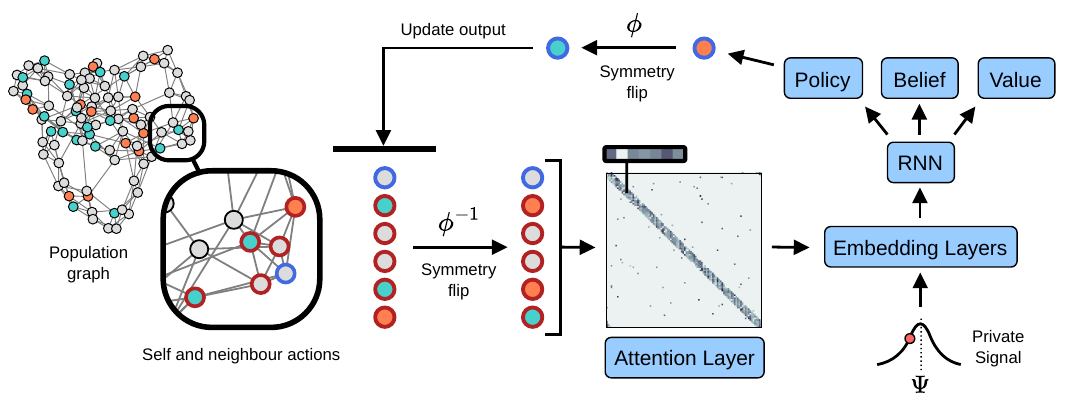}
    \caption{\textbf{Opinion update loop.} Agents receive self and neighbouring guesses, which are subsequently passed through an other-play symmetry operator and learned attention layer. The output is concatenated with agents' private signals and passed into the main body of the architecture, after which the symmetry operation is reversed, yielding the updated guesses of the agents.}
    \label{fig:moneyplot}
\end{figure}

Existing approaches typically rely on hand-crafted local interaction rules which, while remarkably successful~\citep{castellano2009statistical}, limit the expressiveness of agents' behaviours to those intended by the creator of the model. \Ac{marl} offers a promising alternative approach, by allowing agents to learn complex interactions from simple reward functions alone~\citep{foerster2016learning,leibo2017multi,baker2019emergent}.

However, scaling deep \ac{marl} methods to large population sizes is a significant computational challenge~\citep{aymanns2017fake,guo2023reinforcement}. We address these scaling issues by leveraging recent advances in end-to-end GPU-accelerated \ac{rl} with JAX~\citep{jax2018github,flair2024jaxmarl}, enabling training of up to 1000 agents for $10^6$ training steps in 35 minutes on a single NVIDIA A100 GPU.

Specifically, we train agents in our GPU-accelerated \textit{consensus and truth-finding game}, in which they have mixed incentives to make guesses that both match a partially observed true state of the environment and conform to the guesses of their neighbours~\citep{aymanns2017fake,mohseni2021truth}. This captures a simple but fundamental tension between individual evidence and social agreement, while remaining computationally tractable at large population sizes. We release our implementation as open-source software to support future work on scalable MARL-based opinion dynamics.

To accurately model information and consensus dynamics on new unseen ground-truth states, we require that agents do not have a priori preferences over the underlying state. In particular, when agents optimise for conformity, they must not be able to coordinate on a shared initial guess before observing any information from their neighbours. This is a standard \ac{zsc} problem~\citep{hu2020other,hu2021off,muglich2025expected}, which we address by extending the \ac{op} algorithm~\citep{hu2020other} to our general-sum environment. The connection between truth-finding and zero-shot coordination under conformity incentives is exploited in oracle consensus protocols~\citep{sztorc2015truthcoin,risklabs2020uma} but has not previously been explored in multi-agent learning, which could be relevant for large language model (\acs{llm})-based multi-agent systems using agreement as a proxy for correctness~\citep{du2024improving,chen2024reconcile}.

We validate our \ac{marl} framework on a subgraph of the Bluesky social network~\citep{kleppmann2024bluesky} by recovering real-world node importance structures from graph topology alone via learned edge attentions~\citep{liben2003link}. Our method finds that agents acting with 80\% conformity best match importance structures observed in human data. Importantly, at this level of conformity, social media networks yield significantly worse group accuracy and promote the emergence of dishonest actors that guess contrary to their private beliefs~\citep{acemoglu2011bayesian,asch1951effects}. We further show that this behaviour is more pronounced in clustered graphs that exhibit polarised states, such as the U.S. Congress X/Twitter network~\citep{fink2023centrality}. In contrast, we find these effects to be reduced in small networks, such as those observed in the Hadza hunter-gatherer tribe of northern Tanzania~\citep{fedurek2022social,apicella2012social}. Here, conformity can even help to increase the number of accurate outputs when communication is limited, by helping uncertain agents make a guess when the penalty for being wrong exceeds the reward for being correct. Under the assumptions of our model, our findings suggest that the long-timescale evolution of human conformity~\citep{henrich1998evolution,cialdini2004social,deutsch1955study} offers advantages for collective truth finding and agreement in small hunter-gatherer networks, but becomes maladaptive in the context of modern social media platforms.

Our key contributions are:
\begin{enumerate}[leftmargin=2em]
    \item An open-source, scalable and interpretable MARL framework for GPU-accelerated opinion dynamics modelling with JAX, capturing complex emergent dynamics at the scale of 1000 agents with a single NVIDIA A100 GPU.
    \item Extending \ac{zsc} methods to our general-sum setting for accurate modelling of information diffusion with \ac{marl} by preventing the emergence of unrealistic conventions.
    \item Extracting learned attention weights for empirical validation on a real-world social network, finding that strongly conforming agents best match observed node-importance structures.
    \item Demonstrating that conformity in modern social media networks reduces collective accuracy and promotes the emergence of dishonest actors, in contrast to small hunter-gatherer networks.
\end{enumerate}
\section{Related Work}

\citet{wang2022consensus} propose a deep learning approach to opinion dynamics via their consensus boost algorithm, but focus primarily on reward design for rapid consensus, whereas we aim to model human behaviour using simple reward heuristics. \citet{guo2023reinforcement} use a bidirectional \ac{lstm}~\citep{hochreiter1997long} to learn state-conditioned neighbour weights within a linear DeGroot-style update~\citep{degroot1974reaching}, whereas our model adopts a fully neural network-based policy that supports non-linear opinion updates. \citet{aymanns2017fake} use deep learning to study fake news propagation in environments with binary opinions and noisy private signals of some ground-truth state. We build on this setting by introducing a conformity reward and a null action, and scaling to two orders of magnitude more agents. \citet{mohseni2021truth} similarly incorporate a conformity utility into a Bayesian agent model, but do not evaluate its magnitude against real-world data.

\section{Background}
\subsection{Networked Partially Observable Stochastic Games (N-POSG)}
We model opinion dynamics as a \acf{nposg}, given by the 10-tuple
$$(\numagents, \graphspace, \statespace, \{\actspace^i\}_{i=1}^n, \{\obsspace^i\}_{i=1}^n, \transfunc, \{\obsfunc^i\}_{i=1}^n, \rewardfunc, \horizon, \discount).$$
Here, $\numagents \in \mathbb{N}$ is the number of agents, $\graphspace$ is the space of permissible graphs and $\statespace$ is the state space. $\actspace^i$ and $\obsspace^i$ are the local action and observation spaces for agent $i$, such that the joint action and observation spaces are defined by $\actspace \coloneq \prod_{i=1}^n\actspace^i$ and $\obsspace \coloneq \prod_{i=1}^n\obsspace^i$, respectively. $\transfunc$ and $\obsfunc$ define the dynamics of the system,
$$\statett \sim \transfunc(\statett\mid\statet,\actt), \quad \obstt^i \sim \obsfunc^i(\obstt^i\mid\statett,\actt),$$
where $\state \in \statespace$, $\act \in \actspace$ and $\obs^i \in \obsspace^i$.  Rewards are given by $\rewardtt^i = \rewardfunc(\statett,\actt^i)$, $\horizon$ denotes the horizon, where $\state_H$ is a terminal state, and $\gamma \in [0,1]$ is a discount factor.

At each timestep, a graph $\mathcal{G}_t(\agentset, \edgeset) \in \graphspace$ is sampled with a vertex set $\agentset$ of size $|\agentset| = n$ and an edge set $\edgeset$ consisting of $l = |\edgeset|$ directed edges. We denote the out-neighbourhood of $i$ as $N^\mathrm{out}_\grapht(i) = \{j \mid (i,j) \in \edgeset\}$ and the in-neighbourhood as $N^\mathrm{in}_\grapht(i) = \{j \mid (j,i) \in \edgeset\}$, where $(i, j)$ means from agent $i$ to agent $j$. Their respective closed neighbourhoods are $N^\mathrm{out}_\grapht[i] = N^\mathrm{out}_\grapht(i) \cup \{i\}$ and $N^\mathrm{in}_\grapht[i] = N^\mathrm{in}_\grapht(i) \cup \{i\}$.

Each agent $i$ samples a local action $\actt^i$ from a policy $\policy^i(\actt^i | \actobst^i)$, conditioned on their local \ac{aoh} $\actobst^i = (\obs_0^i, \act_0^i, \dots, \obs_{t-1}^i, \act_{t-1}^i, \obst^i)$. The joint policy is defined as $\policy = (\policy^i, \policy^{-i})$, where  $\policy^{-i} = \{\policy^j \mid j  \neq i \}$, and samples actions $\actt$ conditioned on the joint \ac{aoh} $\actobst = (\actobst^1, \dots, \actobst^\numagents)$ with probability $\policy(\actt | \actobst) = \prod_{i=1}^\numagents \policy^i(\actt^i | \actobst^i)$. The distribution of states for a joint policy $\policy$ is given by $\rho^\policy(s)$, and the expected discounted return for each agent $i$ is given by $J^i(\policy) = \mathbb{E}_\policy\left[\sum_{t=0}^{\horizon - 1}\discount^t \rewardtt^i\right]$.

\subsection{Zero-Shot Coordination}
One approach to learning \acp{nposg} is where each agent independently optimises its own policy $\policy^{i*} = \arg\max_{\policy^i} J^i(\policy^i, \policy^{-i})$~\citep{tampuu2017multiagent,dewitt2020ippo}. However, complex environments often give rise to multiple optimal equilibria, which might each rely on different arbitrary conventions. This is problematic in our model because some conventions lead to unrealistic outcomes, such as conforming agents agreeing on the same initial guess to maximise their conformity reward, despite not having observed any signals from their neighbours.

\citet{hu2020other} introduce \ac{zsc} in the context of a \ac{decpomdp}~\citep{bernstein2002complexity}, where agents are trained independently using the same learning algorithm (but different seed) and are then required to coordinate with previously unseen partners at test time. In particular, they introduce the \ac{op} algorithm, which exploits environmental symmetries by replacing the \ac{sp} objective $J_\mathrm{SP}(\policy) = \sum_{i=1}^nJ^i(\policy)$ with the \ac{op} objective $J_\mathrm{OP}(\policy) = \mathbb{E}_{\envsym^1, \dots, \envsym^n \sim \envsymset}\left[J_{\mathrm{SP}}(\envsym^1(\policy^1), \dots, \envsym^\numagents(\policy^\numagents))\right]$, where $\envsymset$ is a set of environmental symmetries. Given that our environment contains a symmetry in the binary choice of guesses, which leaves the underlying \ac{nposg} unchanged, we extend this objective to the general-sum setting such that each agent $i$ maximises $J_{\mathrm{OP}}^i(\policy) = \mathbb{E}_{\envsym^1, \dots, \envsym^n \sim \envsymset}\left[J^i(\envsym^1(\policy^1), \dots, \envsym^\numagents(\policy^\numagents))\right]$. Applying these symmetry operations randomly to the policy prevents agents from arbitrarily preferring one guess over another unless there is an explicit advantage to doing so based on their local observations.

\subsection{Opinion Dynamics and Information Aggregation}
The field of opinion dynamics studies how exchanges of local opinions can give rise to emergent global states. Many foundational works use continuous variables as opinions to study dynamics with weighted  averaging~\citep{french1956formal,degroot1974reaching}, stubborn agents~\citep{friedkin1990social} and bounded confidence~\citep{deffuant2000mixing,hegselmann2002opinion}, while others analyse discrete variable models analogous to spin states in statistical physics~\citep{holley1975ergodic,galam1982sociophysics}. Recently, \ac{abm} approaches have become popular~\citep{epstein1996growing,flache2017models}, allowing for more complex local interaction behaviours.

Another important class of models uses Bayesian agents to study information aggregation of some underlying state~\citep{bikhchandani1992theory,banerjee1992simple,gale2003bayesian,acemoglu2011bayesian}. Such models have been used to show how polarised states can emerge from rational agents~\citep{oconnor2018scientific,madsen2018large} and how the addition of conformity utilities can affect consensus dynamics~\citep{mohseni2021truth}.

A third type of model looks at \ac{rl} agents. Such models have been developed to demonstrate the emergence of social norms~\citep{airiau2014emergence,yu2014collective,yu2016modelling}; discuss mechanisms for meta-stable polarised states via social reinforcement~\citep{banisch2019opinion,meylahn2024social}; and explore consensus reaching algorithms for optimal decision making~\citep{shen2025consensus,wang2022consensus}. Recently, some deep \ac{marl} approaches study social network dynamics with small population sizes~\citep{aymanns2017fake,guo2023reinforcement}.
\begin{figure}[t]
    \centering
    \includegraphics[width=\textwidth]{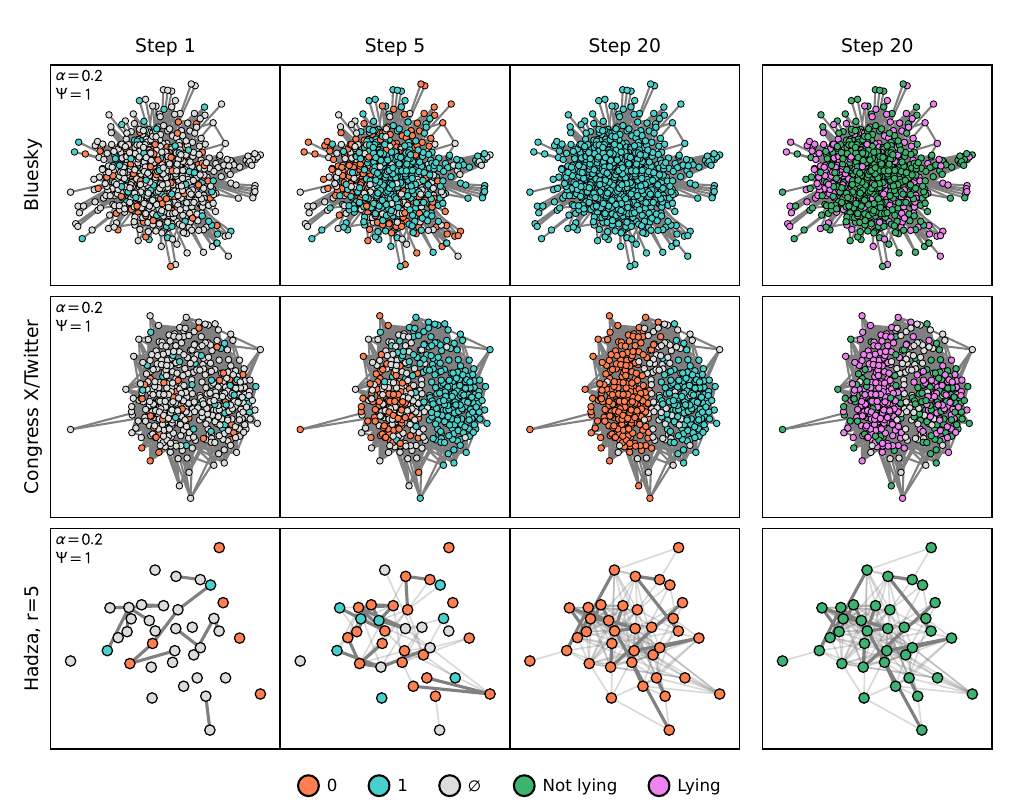}
    \caption{\textbf{Episode dynamics.} Example episodes for reward weighting $\weight = 0.2$, selected to demonstrate interesting outcomes. \textbf{Top row:} Bluesky network with many dishonest agents despite the population guessing correctly. \textbf{Middle row:} Congress network with a polarised state and a majority of lying agents in the left cluster, despite signalling incorrectly. \textbf{Bottom row:} Hadza network ($r=5$) with no lying agents despite the population guessing incorrectly. In general, the Hadza network exhibits fewer lying agents than the social media networks.}
    \label{fig:dynamics}
\end{figure}

\section{Methods} \label{sec:methods}

\subsection{Our Consensus and Truth-Finding Game} \label{sec:consensus_game}
In this game, agents are tasked with the goal of identifying a binary ground-truth value via communication with neighbouring agents in a social network. Before an episode begins, a ground-truth value $\gtruth \in \{0,1\}$ is uniformly sampled. For all $0\leq t \leq H$, we sample a graph $\grapht(\agentset, \edgeset(t))$ with a set of agents $\agentset$ and a set of interactions $\edgeset(t)$. This graph can be static or dynamic. The discrete action space of each agent is $\actspace^i = \{0, 1, \nullact\}$, where 0 and 1 are guesses of $\gtruth$ and $\nullact$ is a null action. States are represented by tuples $\statet = (\grapht, \gtruth, t, \act_{t-1})$ and observations consist of two parts:
\begin{itemize}[leftmargin=2em]
    \item For each $i$, with probability $p_\mathrm{signal}$, $\obs^i_{(1)}$ is independently sampled from a Gaussian distribution $\mathcal{N}(\gtruth,\std^2)$. Otherwise, it is set to 0.5. $\obs^i_{(1)}$ is held constant for all $0\leq t\leq \horizon$.
    \item $\obs^i_{(2)}(t) = (\act^j_{t-1})_{j\in N^\mathrm{out}_\grapht[i]}$, a tuple showing the previous actions of all neighbouring agents $j$ being paid attention to, including $i$'s own action. At $t=0$, we initialise $\act^j_{-1} = \nullact$ for all $j$.
\end{itemize}

Combining these gives us $\obst^i = (\obs^i_{(1)}, \obs^i_{(2)}(t))$. At each timestep, agents receive a reward $\rewardtt^i = \weight \accuracy^i_t + (1 - \weight) \conformity^i_t$, where $\weight \in [0,1]$,
$$
\begin{aligned}
\accuracy^i_t &=
\begin{cases}
    0.1 & \text{if } \actt^i = \gtruth, \\
    -0.2 & \text{if } \actt^i = 1 - \gtruth, \\
    0 & \text{if } \actt^i = \nullact,
\end{cases}
\qquad
\conformity^i_t &=
\frac{1}{|N^\mathrm{out}_\grapht(i)|}\sum_{j \in N^\mathrm{out}_\grapht(i)}
\begin{cases}
    0.1 & \text{if } \actt^i = \actt^j \neq \nullact, \\
    -0.2 & \text{if } \actt^i \notin \{\actt^j, \nullact\}, \\
    0 & \text{otherwise.}
\end{cases}
\end{aligned}
$$
Thus, $\accuracy$ and $\conformity$ incentivise accuracy and conformity, respectively, with $\weight$ controlling the weighting. When $\weight = 0$, agents are fully conforming and when $\weight = 1$, agents are fully truth-seeking. While simple, this reward function yields more expressive learned policies than basic hand-crafted baselines (see \Cref{app:expressivity}). We use an asymmetric reward to incentivise the null action over uncertain guessing by ensuring the latter has a lower expected return (see \Cref{app:model_robustness}). 

\subsection{Model Architecture} \label{sec:model_architecture}
A diagram of the model is shown in \Cref{fig:moneyplot}. We pass the neighbour observations through an attention layer~\citep{vaswani2017attention}, masked by $\grapht$ to aggregate information according to learned attention weights, adding learnable absolute positional embeddings of agent IDs to distinguish between individuals. We adopt the standard practice of parameter sharing~\citep{gupta2017cooperative}, including a third input in the form of a binary-encoded agent ID to allow for heterogeneity in agent policies. The private signal, $\obs^i_{(1)}$, attention-weighted encodings of past neighbour actions, $\mathrm{attn}(\obs^i_{(2)}(t))$, and binary-encoded agent ID, $\mathrm{bin}(i)$, are concatenated, passed through two fully-connected embedding layers, a \ac{gru}, two more fully-connected layers, and finally into an action distribution head, a belief head, and a value function head. All fully-connected layers and the \ac{gru} hidden state have a size of 50 and use layer normalisation. See \Cref{app:ablations} for more on agent IDs and memory.

\subsection{Training Algorithm} \label{sec:training_algorithm}
We train agents with \ac{op} and \ac{ippo}~\citep{dewitt2020ippo}, where each agent independently updates its policy via \acs{ppo}~\citep{schulman2017proximal}. Updates are performed at the end of each trajectory to prevent feedback on $\gtruth$ during episodes, as agents will not know this value during evaluation.

For the belief head, we freeze all weights except the belief layer and train with a supervised \ac{mse} between the belief output and $\gtruth$ over the episode timesteps. This is similar to a truth-seeking agent ($\weight = 1$) that acts myopically ($\discount = 0$) and tells us what an agent's current belief of $\gtruth$ is from its hidden state.

When computing attentions, we either evaluate the full attention matrix, masked by the graph adjacency matrix, or calculate only the required attentions via a centralised edge list of sent and received signals. The first method scales as $O(n)$ for a single agent and $O(n^2)$ across the population, whereas the second always scales as $O(l+n)$. When training large populations, $l+n \ll n^2$, so we use a centralised edge list. However, when training only a small subset of agents, for example if we want to test unilateral deviation of an agent for \ac{ne} convergence (\Cref{app:nash}), $l+n \gg n$, since graphs typically contain many more edges than agents. In this case, masking is more efficient. We therefore adopt the implementation that is optimal for the scenario being considered.

The consensus and truth-finding game formalised in \cref{sec:consensus_game} contains two important environmental symmetries, $\envsymset = \{\identity, \envsym'\}$, where $\identity$ leaves the environment unchanged and $\envsym' = (\envsym'_\statespace, \envsym'_\actspace, \envsym'_\obsspace)$, with
$$
\begin{aligned}
\envsym'_\actspace(\act^i) &=
\begin{cases}
    1 - \act^i & \text{if } \act^i \neq \nullact, \\
    \act^i & \text{if } \act^i = \nullact,  
\end{cases}
\qquad
\begin{aligned}
\envsym'_\statespace(\grapht, \gtruth, t, \act_{t-1})
&= (\grapht, 1 - \gtruth, t, \envsym'_\actspace(\act_{t-1})), \\[0.5em]
\envsym'_\obsspace(\obs^i_{(1)}, (\act_{t-1}^j)_{j \in N^\mathrm{out}_\grapht[i]})
&= (1 - \obs^i_{(1)}, (\envsym'_\actspace(\act_{t-1}^j))_{j \in N^\mathrm{out}_\grapht[i]}).
\end{aligned}
\end{aligned}
$$
Note that $\envsym'_\actspace$ and $\envsym'_\obsspace$ act element-wise on $\actspace$ and $\obsspace$, respectively. More intuitively, the relabelling of 0s and 1s is a symmetry of the environment, requiring us to also flip the private signal observations. At the start of every episode, we randomly assign a symmetry to each agent, applying this to its observation and action at each timestep. Thus, agents cannot coordinate on an initial guess and can only reach an agreement after multiple rounds of signalling. Note that we use \ac{op} for both training and evaluation, as restricting this just to training can result in unintended conventions during evaluation (see \Cref{app:ablations}).

When $\weight = 0$, a new symmetry appears, as the expected return is no longer conditioned on the underlying truth state. For example, we get an unwanted but \ac{op}-compatible convention, whereby agents learn to use their private signals as an anchor to coordinate towards or away from (since both give the same expected return). For this edge case, we simply set $p_\mathrm{signal} = 0$, so that agents have no information about the ground truth to form this convention on.

\subsection{Edge Weight Extraction and Empirical Validation} \label{sec:edge_weight_extraction}
Given the abstract nature of our model, comparison with real-world data is generally challenging. However, one option is to use edge weights of the social graph. Since attention weights depend on both neighbouring signals and agent IDs, edge weights cannot, in general, be uniquely inferred. For example, we could take the expected weights under the state distribution $\rho^\policy(s)$, or set all actions equal to remove the dependence on neighbouring signals. We opt for the former, approximating the expectation using Monte Carlo trajectory sampling, though this choice has little effect on our results (see \Cref{app:extracting_attention_weights}).

A straightforward way to compare learned edge weights against empirical data is with simple similarity measures such as \ac{mse} or cosine similarity. However, these metrics are largely insensitive to distributional differences beyond the mean (\Cref{app:validation_metrics}). We therefore analyse a node-level metric, which we term the \ac{pi}, for which we can construct distributions conditioned on node in-degree for more meaningful comparisons. We define the \ac{pi} centrality measure by the average importance received from an agent's closed in-neighbourhood, where importance is calculated by multiplying the attention weighting by the neighbouring (or self) agent's closed neighbourhood out-degree, given by
\begin{equation}
C_\mathrm{PI}(i) = \frac{1}{|N^\mathrm{in}_\graph[i]|}\sum_{j\in N^\mathrm{in}_\graph[i]} |N^\mathrm{out}_\graph[j]|w_{ji}.
\label{eq:pi}
\end{equation}
This measure is designed such that, if every agent divides their attention uniformly, every agent's $C_\mathrm{PI}$ is exactly 1. A score greater than 1 means that, on average, neighbouring agents place more attention on agent $i$ than on their other neighbours, with the reverse holding for a score of less than 1. Multiplying by the neighbours' out-degrees removes a graph-induced bias that gives central nodes high scores under standard attention-based centrality measures, since we care about the \textit{perceived} importance from the local perspective of neighbouring agents. For a more in-depth discussion on why we choose this centrality measure, see \Cref{app:centrality_measures}.

\begin{figure}[t]
    \centering
    \includegraphics[width=\linewidth]{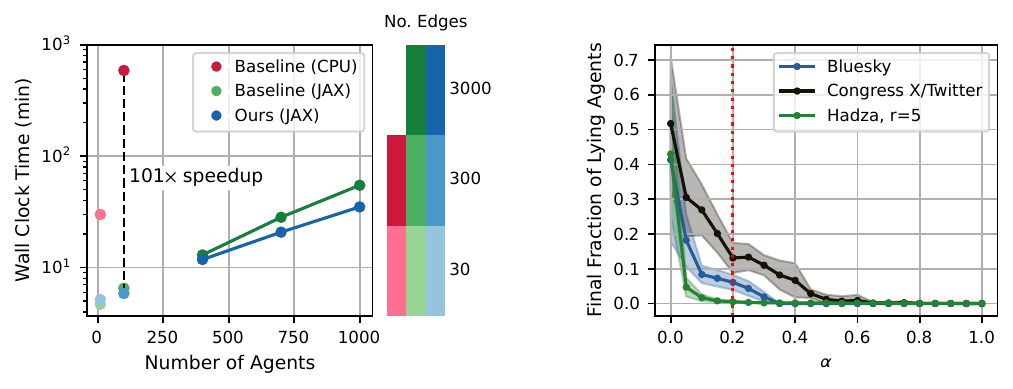}
    \caption{\textbf{Left:} Wall clock times for $10^6$ training steps with 20 parallel environments. The JAX implementations trained on a single GPU scale orders of magnitude better than the baseline model trained on a CPU. The improved parameter efficiency of our model leads to additional performance gains. \textbf{Right:} Fraction of lying agents that output the opposite non-null guess to the output of their belief head. Large social media networks promote dishonest agents for conforming populations.}
    \label{fig:scaling}
\end{figure}

\section{Experimental Details} \label{sec:experimental_details}
\subsection{Datasets}
Direct validation of network models on real data is challenging, as true underlying interaction processes are only partially observed.
We generate a 1000-node dataset of the Bluesky machine learning community, with edge attentions computed from user-to-user likes, yielding a network whose interactions align with our modelling assumptions. We further study a 475-node X/Twitter network of the 117th U.S. Congress~\citep{fink2023centrality} and a 37-node GPS-derived proximity dataset of the Hadza hunter-gatherer tribe from northern Tanzania~\citep{fedurek2022social}, where interactions are defined as spatial separations of less than two metres. (See \Cref{app:datasets} for more details.)

\subsection{Modelling the Dynamics of the Hadza Network}
While the Bluesky and Congress networks remain static, modelling the Hadza network's dynamic edges introduces additional complexity. We model edge dynamics using a Markov process conditioned on steady-state interaction probabilities from the aforementioned GPS-derived dataset. This leaves a single free parameter, $r$, which controls the rate of switching between interacting and non-interacting states. To discretise the process, we define an edge to exist at time $t$ if at least one interaction occurred during $[t-1, t]$. Consequently, $r$ can also be interpreted as controlling the effective time resolution of the model. Thus, larger values of $r$ correspond to increasing the number of neighbour interactions before updating one's guess. We study $r=5$, as interactions are frequent enough for convergence to consensus at later timesteps while early timesteps highlight performance differences between truth-seeking and conforming populations. See \Cref{app:hadza_tribe} for full details of the Markov model and this choice of $r$.

\subsection{Controlling Signal Density}
Since the Bluesky and Congress X/Twitter networks have significantly higher edge densities than the Hadza network, agents receive proportionally more information from initial neighbouring guesses. To control for this, we introduce additional sparsity in the private signals by reducing $p_\mathrm{signal}$ to match the overall neighbourhood signal density of the Hadza network with $r=5$. For the Hadza network, we keep $p_\mathrm{signal} = 1$. For the Bluesky and Congress X/Twitter networks, we set $p_\mathrm{signal}$ to 0.14 and 0.08, respectively, decaying these values from unity during training for improved learning stability. See \Cref{app:signal_densities} for more details on signal densities.

\subsection{Implementation Details}
We use episodes of length $H=20$ across 20 parallel environments. We vary reward weightings $\weight \in [0, 1]$ in increments of 0.05 and set $\std = 1.7$ for the private signals. This provides signals that are weak enough to encourage extended communication, while remaining sufficiently informative for learning (see \Cref{app:model_robustness}). We train 10 random seeds, one of which collapses to the null action for multiple values of $\weight$. We evaluate the remaining 9 seeds on 200 episodes generated from an unseen random seed. Agent populations do not converge to Nash equilibrium, but this does not invalidate the model (see \Cref{app:nash}). Hyperparameters and training curves are given in \Cref{app:hyperparameters}. All error bars indicate the standard error of the mean unless otherwise specified.

\section{Results and Analysis} \label{sec:results}

\begin{figure}[t]
    \centering
    \includegraphics[width=\linewidth]{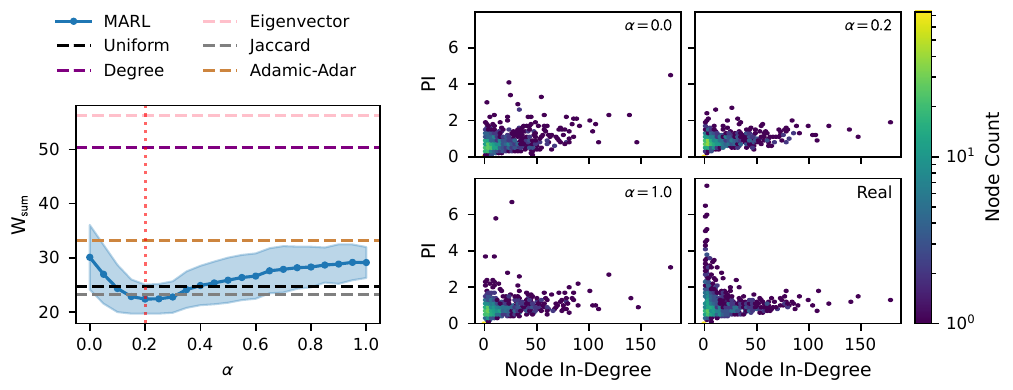}
    \caption{\textbf{Validation of Bluesky network.} \textbf{Left:} Sum of Wasserstein distances for each node in-degree. The dotted red line indicates the minimum of the \ac{marl} curve. \textbf{Right grid:} Hexagonal histograms showing \ac{pi} distributions as a function of node in-degree, comparing \ac{marl} prediction at $\weight =$ 0, 0.2, and 1 with the real data. $\weight = 0.2$ accurately captures the shape of the real data.}
    \label{fig:validation}
\end{figure}

\subsection{Scaling}
We train our model for $10^6$ training steps on a single NVIDIA A100 GPU comparing wall clock times with the CPU-based baseline model from \citet{aymanns2017fake} and a reimplementation of this baseline in a GPU-accelerated JAX pipeline, shown in \Cref{fig:scaling}. For a 100-agent population, the JAX pipeline is 101$\times$ faster than the CPU baseline and the population scaling is orders of magnitude more tractable. This enables us to train 1000 agents in 35 minutes. Our model is also more parameter efficient than the baseline, scaling with population size $n$ at a rate of $8n$ compared to $768n$ (\Cref{app:scaling}). When $n=1000$, the parameter counts are 35\,102 and 809\,858, respectively.

\subsection{Agent Importance Prediction}\label{sec:pi}
We validate our model against the Bluesky network dataset by comparing the distributions of \ac{pi} scores for different node in-degrees. As the real data has no self-attentions, we remove the learned self-attentions from our model and renormalise for a fair comparison. \Cref{fig:validation} shows the sum of the Wasserstein distances between predicted and empirical \ac{pi} scores across node in-degree, denoted $W_\mathrm{sum}$. The learned weights are competitive with the strongest heuristic baselines (\Cref{app:edge_weight_baselines}), and best match the human data when $\weight = 0.2$, indicating that agents with strong conformity most closely approximate humans in our dataset. Interestingly, the real data exhibits a spike in high \ac{pi} scores among low in-degree nodes, which our model fails to capture. This effect is driven by high out-degree nodes distributing their attentions highly non-uniformly, thereby inflating the average scores of a subset of low in-degree nodes. Our model also struggles to predict structures in the Congress network, likely due to additional incentives that are not included in our model, such as political influence (see \Cref{app:validating_congress}). However, we still include this graph in our analyses as an example of emergent dynamics on a clustered topology given our chosen incentives.

\begin{figure}[t]
    \centering
    \includegraphics[width=\linewidth]{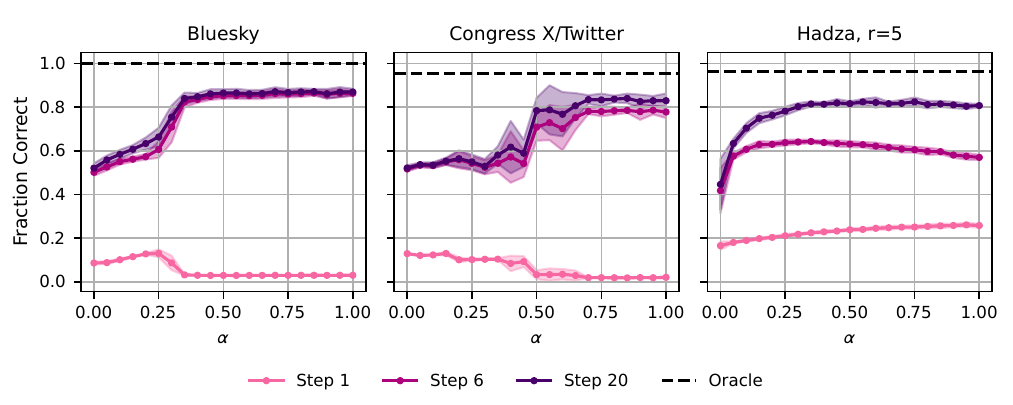}
    \caption{Fraction of agents whose output matches the ground truth at various time steps for all reward weightings $\weight$ on three distinct graphs. The black dashed line is an oracle that acts Bayes-optimally in isolation given $\lfloor np_\mathrm{signal}\rfloor$ independently sampled private signals. Finite conformity is advantageous in the Hadza network for intermediate timesteps.}
    \label{fig:accuracy}
\end{figure}

\subsection{From Hunter-Gatherer to Social Media}\label{sec:accuracy}
\Cref{fig:accuracy} shows how the accuracy of a population depends on the reward weighting $\alpha$ and network structure. For comparison, we include an oracle which calculates the accuracy of a Bayes-optimal guess given $\lfloor np_\mathrm{signal}\rfloor$ independently sampled private signals. This is equivalent to the best possible performance of a fully truth-seeking population under perfect communication.

In the Hadza network, agents with finite conformity achieve more correct guesses during intermediate steps of an episode, an advantage that becomes more pronounced with less frequent communication (see \Cref{app:hadza_tribe}). Although truth-seeking agents tend to be accurate when they do guess (see \Cref{app:notions_of_accuracy}), the asymmetric rewards incentivise them to choose the null action when uncertain. Conformity incentives, however, can help to mitigate this effect by offering a reliable payoff for non-null actions. This suggests that conformity may be advantageous for collective agreement in small, dynamic hunter–gatherer networks, where communication is restricted and decisions are time-sensitive.

In contrast, conformity appears maladaptive in social media networks. The Bluesky and Congress networks exhibit sharp transitions in accuracy around $\weight=0.3$ and $\weight = 0.45$, respectively, beyond which increased conformity yields worse accuracy than the Hadza network. This is notable given that the Bluesky network receives roughly four times as many total private signals compared to the Hadza network. Additionally, clustered graphs such as the Congress network become more susceptible to polarised states (see \Cref{fig:dynamics}). One possible explanation for these observations is that, in conformity-driven populations, agents without private signals guess immediately rather than waiting. In larger networks, the resulting noise may drown out truth-conveying signals, making coordination more difficult (see \Cref{app:interpreting_learned_policies}).  More generally, we find that the larger sizes and specific topologies of the social media networks both contribute to worse overall accuracy (see \Cref{app:size_and_topology}).

\subsection{Dishonesty in Conforming Populations}\label{sec:dishonesty}
\Cref{fig:scaling} (right) demonstrates how conformity promotes dishonest behaviour, with agents acting contrary to their beliefs in order to agree with their neighbours. This can occur when an agent's posterior belief in the ground truth exceeds the population's accuracy (\Cref{app:belief_head}). As conformity increases, agents' actions become progressively less correlated with the ground truth, with $\alpha = 0$ corresponding to the limiting case of zero correlation.

We find that the fraction of dishonest agents is greater in the larger social media networks than in the smaller Hadza network, potentially because the increased difficulty of coordination in larger networks discourages agents from seeking the ground truth. \Cref{fig:dynamics} illustrates an extreme case in the Congress network, where the majority of agents in the left cluster act dishonestly despite correctly identifying the ground truth. Notably, these results arise in the absence of malicious actors deliberately spreading misinformation, suggesting that misinformation can emerge even among agents with identical incentives.

\section{Conclusions} \label{sec:conclusions}

Our primary contribution is a scalable and interpretable GPU-accelerated MARL framework for modelling opinion dynamics in populations of up to 1000 agents on a single GPU. We extend the \ac{op} algorithm to our general-sum setting, accurately modelling information diffusion by preventing the emergence of unrealistic conventions. We extract learned attention weights to compare the model with real-world node-importance structures, finding the best match in a Bluesky network dataset for agents with strong conformity ($\weight = 0.2$).

We further use the framework to investigate how conformity and network structure affect collective behaviour. We show that conformity can improve collective accuracy in a small hunter-gatherer network, but reduces accuracy and promotes dishonest behaviour in the larger social media networks studied. These findings suggest a possible evolutionary mismatch, whereby human conformity heuristics may be poorly suited to modern social media environments, potentially increasing susceptibility to misinformation.

However, our model does not account for factors such as recommendation systems, bounded human attention, influence-seeking behaviour, or heterogeneity in agent incentives. Incorporating these factors into the framework presents a promising direction for future research in \ac{marl} for opinion dynamics modelling.

\phantomsection
\section*{Societal Impacts} \label{sec:impact_statement}
The goal of this paper is to advance the understanding of opinion dynamics in real-world social networks using \ac{marl}. This work has the potential for both positive and negative societal impacts. In particular, improved models of opinion formation can help identify mechanisms that contribute to distrust, polarisation, and the spread of misinformation, providing insight into potential intervention strategies.

We also acknowledge the risk that results presented in this paper could be misinterpreted or taken out of context. For example, overly broad conclusions could be drawn about the behaviours or intentions of specific social or political actors. We emphasise that our model is intentionally simplified and does not capture many of the complex contextual and psychological factors that influence real human interactions. We have additionally anonymised the users in our generated Bluesky dataset.

Overall, we believe that the potential benefits of this research outweigh the associated risks, provided the results are interpreted with appropriate caution.
\section*{Acknowledgements}
The authors would like to thank Eugene Vinitsky for suggesting the creation of the Bluesky dataset. They would also like to thank Zilin Wang, Christoph Aymanns, Bidipta Sarkar, Dylan Cope, Johannes Forkel, Ravi Hammond, Shashank Reddy and Aavash Subedi for helpful discussions.

Compute for this project was graciously provided by JASMIN, the UK’s collaborative data analysis environment (\url{https://www.jasmin.ac.uk}). LS was supported by the Intelligent Earth CDT, funded by UKRI grant EP/Y030907/1. BK was supported by Schmidt Sciences. ST was supported by the EPSRC Centre for Doctoral Training in Autonomous Intelligent Machines \& Systems (grant EP/S024050/1). JF was partially supported by UKRI grant EP/Y028481/1 (originally selected for funding by the ERC). JF also received support from the JPMC Research Award and the Amazon Research Award.

\bibliography{references}
\bibliographystyle{plainnat}


\appendix

\section{Datasets} \label[appendix]{app:datasets}

\subsection{Bluesky Machine Learning Community}\label[appendix]{app:bluesky_ml}
We construct a dataset of the Bluesky machine learning community, with edge weights designed to reflect the behaviours relevant to our model. To construct the graph, we start with three anonymised users and do a \ac{bfs} crawl, where for each next user we take the 50 most recent posts (or fewer if the user has fewer than 50) and find all the likers of these posts, capped at the 100 most recent likers for each post. From here, we sample 30 new next users in the \ac{bfs} crawl, conditioned on having posted on the platform in the past 30 days. By constructing the graph in this way, we hope to capture the active members of the community and avoid including bot accounts, which we found to be included when generating the graph based on raw follows.

After completing the \ac{bfs} crawl, we perform an exhaustive search of all inter-account likes and calculate a ``likes fraction'' score for each directed edge, where this score is simply the number of likes user $i$ sends to user $j$ divided by the number of collected posts from user $j$. Finally, to get the attention weights we normalise the outgoing ``likes fraction'' scores for each agent, so that the outgoing attentions sum to 1. The final graph consists of 1000 nodes and 14\,559 directed and weighted edges.

This dataset was created from the public Bluesky API on 5 November 2025.

\subsection{U.S. Congress X/Twitter}
\begin{figure}[h]
    \centering
    \includegraphics[width=0.35\linewidth]{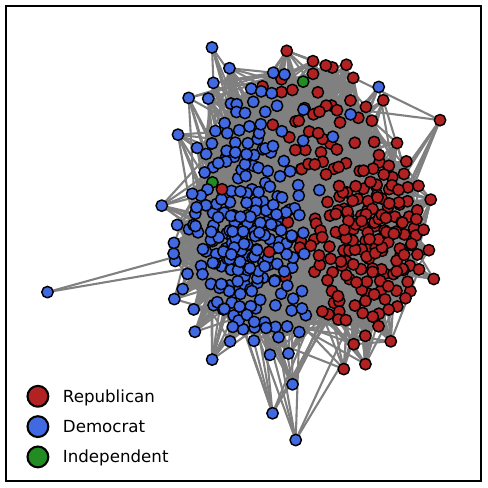}
    \caption{117th U.S. Congress X/Twitter network classified by party association. Username to party maps were parsed with~\citep{frac2021twitter}.}
    \label{fig:congress_parties}
\end{figure}

\label[appendix]{app:congress_twitter}
We use a dataset of the 117th U.S. Congress X/Twitter network consisting of interactions between members of the 117th U.S. Congress on X/Twitter between 9 February 2022 and 9 June 2022~\citep{fink2023centrality}. The network consists of 475 nodes and 13\,289 directed and weighted edges. The network is mainly split into two distinct subgraphs, corresponding to members associated with either the Republican or Democratic parties (\Cref{fig:congress_parties}), with fewer connections between opposing party members. Each weight associated with a directed edge from member $i$ to $j$ indicates the fraction of times that member $i$ retweeted, quote retweeted, replied to, or mentioned $j$'s tweets, summed and then divided by the total number of tweets member $j$ issued during the time frame to give an empirical probability of any of member $j$'s tweets being reacted to by $i$. Only members with at least 100 tweets in the time frame were included. 

\subsection{Hadza Tribe}\label[appendix]{app:hadza_tribe}
We use a GPS-derived dataset of the Hadza tribe in northern Tanzania~\citep{fedurek2022social}, consisting of values indicating the fraction of the day that individual members of the tribe spend within two metres of one another. Specifically, we opt for the ``camp 2 -- out of camp'' dataset. This dataset originally contains 38 individuals, from which we remove the single individual with no recorded interactions, leaving a network of 37 agents. As this network is naturally highly dynamic on a minute-to-minute basis, we use a Markov process to model interactions during each episode.

To derive the interaction dynamics, we start with the continuous time \ac{ode},
\begin{align}
    \frac{\mathrm{d} \PrIt}{\mathrm{d}t} = a(1-\PrIt) - b\PrIt,
\end{align}
where $a$ is the rate of switching from not interacting ($\notinteract$) to interacting ($\interact$), and $b$ is the rate of switching from $\interact$ to $\notinteract$. Solving this first-order equation gives us,
\begin{align}
    \PrIt = \PrIo \mathrm{e}^{-(a+b)t} + \frac{a}{a+b}\left(1-\mathrm{e}^{-(a+b)t}\right),
\end{align}
which in the steady state limit yields
\begin{align}
    q = \lim_{t \to \infty} \PrIt = \frac{a}{a+b}.
\end{align}
If we start from an interacting state where $\PrIo = 1$ or a non-interaction state where $\PrIo = 0$ and let $r = a + b$, we get
\begin{align}
    &\Pr(\interact_{t + \Delta t} \mid \interact_t) = q + (1 - q)\mathrm{e}^{-r\Delta t}, \\
    &\Pr(\interact_{t + \Delta t} \mid \notinteract_t) = q(1 - \mathrm{e}^{-r\Delta t}).
\end{align}
In order to discretise these solutions, we pose the question: given the state at $t$ and $t+\Delta t$, what is the probability of an interaction having occurred in the time interval $\Delta t$? The cases where either the start or end state is interacting are trivial, giving a probability of 1. This leaves us with the non-trivial case where both the start and end states are non-interacting.
We first find the probability of no single interaction given an initial non-interacting state $\Pr(\notinteract_{(t, t+\Delta t)} \mid \notinteract_t)$. Consider a subdivision of $\Delta t$ into $m$ partitions, each of length $\delta t = \frac{\Delta t}{m}$. In the limit as $m \to \infty$, we have
\begin{align}
    \Pr(\notinteract_{(t, t + \Delta t)} \mid \notinteract_t) &= \lim_{m\to\infty}\left(1 - q\left(1-\mathrm{e}^{-r\delta t}\right)\right)^m \\
    &= \lim_{m\to\infty}\left(1 - q\left(1 - 1 + r\delta t\right)\right)^m \\
    &= \lim_{m\to\infty}\left(1-\frac{qr\Delta t}{m}\right)^m \\
    &= \mathrm{e}^{-rq\Delta t}.
\end{align}
where the final equality uses the following well-known limit, 
\begin{equation*}
\lim_{n\to\infty} \left(1+\frac{x}{n}\right)^{n} = e^{x}, \quad \forall x \in \mathbb{R}.
\end{equation*}
Then we have
\begin{align}
    \Pr(\interact_{(t, t+\Delta t)} \mid \notinteract_{t+\Delta t}, \notinteract_t) &= \frac{\Pr(\interact_{(t, t+\Delta t)} \cap \notinteract_{t+\Delta t} \mid \notinteract_t)}{\Pr(\notinteract_{t + \Delta t} \mid \notinteract_t)} \\
    &= \frac{\Pr(\notinteract_{t+\Delta t} \mid \notinteract_t) - \Pr(\notinteract_{t+\Delta t} \cap \notinteract_{(t, t+\Delta t)} \mid \notinteract_t)}{\Pr(\notinteract_{t + \Delta t} \mid \notinteract_t)} \\
    &= 1 - \frac{\mathrm{e}^{-rq\Delta t}}{1 - q\left(1 - \mathrm{e}^{-r\Delta t}\right)}.
\end{align}

\begin{figure}[h]
    \centering
    \includegraphics[width=\linewidth]{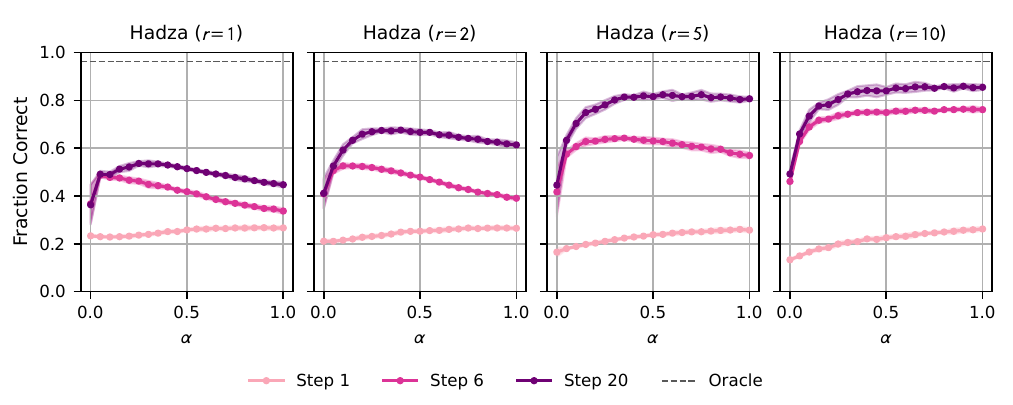}
    \caption{Effect of the interaction frequency $r$ on the fraction of agents whose output matches the ground truth at various time steps for all reward weightings $\weight$ for the 37-node Hadza network. The black dashed line is an oracle that acts Bayes-optimally in isolation given $\lfloor np_\mathrm{signal}\rfloor$ independently sampled private signals. Increasing the frequency of interaction improves final accuracy.}
    \label{fig:hadza_r_comparison}
\end{figure}

Since $q$ is determined by the dataset, this leaves us with one free parameter $r$, which can equivalently be thought of as controlling the rate of switching or setting the time resolution of the discretisation. \Cref{fig:hadza_r_comparison} shows the effect of changing $r$ on the fraction of correct guesses in the population. We clearly see that the final accuracy improves with increasing interaction frequency. We choose $r=5$ to compare against, as the interactions are frequent enough for fully truth-seeking agents to converge on a guess, while still highlighting the advantage that conformity can have during earlier time steps. This gives an average of 44.1 edges per time step or 1.19 average edges per time step per agent.

\section{Scaling}\label[appendix]{app:scaling}
For a fair comparison between the \acs{iql} baseline and \acs{ppo} algorithms, we train with the same number of parallel environments (20) and gradient updates. Additionally, we restricted the action space to two possible actions $\{0,1\}$ to maintain consistency with the original \acs{iql} baseline implementation. However, the parameter counts are different, as our neural network architecture scales more efficiently than the baseline. With $n$ as the number of agents, $a$ the number of actions, $d$ the attention embedding dimension, $f$ and $h$ the feed-forward and GRU hidden size ($f=h$ in our case), the parameter counts are given by,
\begin{equation}
\begin{aligned}
    |\theta_{\text{ours}}|
    &=
    \underbrace{\left(dn + 3d^2 + 7d\right)}_{\text{attention layer}}
    \;+\;
    \underbrace{\left(3fh + 3h^2 + 4h\right)}_{\text{GRU}}
    \;+\;
    \underbrace{\left((1+d+\lceil \log_2 n\rceil)f + f^2 + 4f\right)}_{\text{dense layers}}
    \\[4pt]
    &\quad+\;
    \underbrace{\left(h^2 + (a+2)h + a\right)}_{\text{policy head}}
    \;+\;
    \underbrace{\left(hf + 2f + h + 1\right)}_{\text{value head}}
    \;+\;
    \underbrace{\left(hf + 2f + h + 1\right)}_{\text{belief head}}
    \\[4pt]
    &=
    3d^2 + 7d + 10h^2 + (d + a + 17)h + a + 2 + dn + h\lceil \log_2 n \rceil,
\end{aligned}
\end{equation}
and 
\begin{equation}
\begin{aligned}
|\theta_{\text{baseline}}|
&=
\underbrace{\left(3(a+2)hn + 3h + 3h^2 + 4h\right)}_{\text{GRU 1}}
\;+\;
\underbrace{\left(6h^2 + 4h\right)}_{\text{GRU 2}}
\;+\;
\underbrace{\left(h^2 + h\right)}_{\text{dense}}
\;+\;
\underbrace{\left(ah + a\right)}_{\text{output}}
\\[4pt]
&=10h^2 + (a+12)h + a + 3h(a+2)n.
\end{aligned}
\end{equation}

Setting $h=50$ and $64$ for our model and the baseline model, respectively, $d=8$, and $a=2$, we obtain the simple expressions,
\begin{equation}
    |\theta_{\text{ours}}| = 26\,602 + 8n + 50\lceil \log_2 n \rceil,
\end{equation}
\begin{equation}
    |\theta_{\text{baseline}}| = 41\,858 + 768n.
\end{equation}
We clearly see that our model size scales slower with the number of agents, resulting in 23$\times$ fewer parameters when $n=1000$.
\section{Expressivity Beyond Simple Baselines} \label[appendix]{app:expressivity}

\begin{figure}[h]
    \centering
    \includegraphics[width=0.9\linewidth]{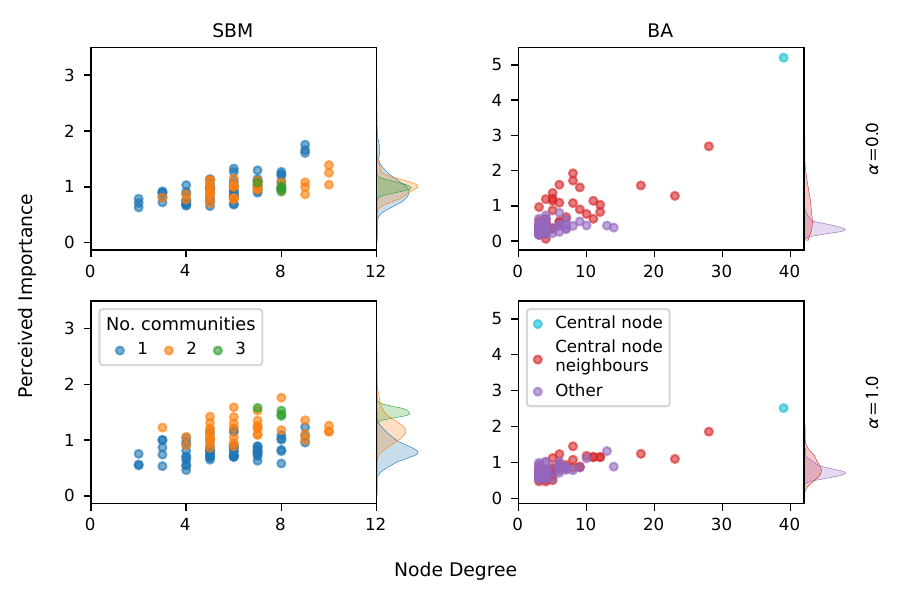}
    \caption{Perceived importance for an \acs{sbm} and \acs{ba} graph with 100 agents. Truth-seeking agents ($\weight = 1$) on the \acs{sbm} graph place high importance on agents that bridge multiple clusters, while the conforming population ($\alpha=0$) on the \acs{ba} graph finds an efficient coordination structure with information flowing from the central node.}
    \label{fig:100_agents}
\end{figure}

To demonstrate the expressivity of our model, we run our model on two generated graphs of size 100. The first graph is a \ac{sbm} with five clusters and probabilities of within-cluster and inter-cluster edges of 0.33 and 0.005, respectively, and the second graph is a \ac{ba} graph with attachment parameter $m=3$. From \Cref{fig:100_agents} we see that on the \ac{sbm} graph, truth-seeking agents ($\weight = 1$) learn to place proportionally greater importance on neighbours that bridge multiple clusters, as these nodes can effectively transfer information between clusters. We also see that conforming agents ($\weight = 0$) on the \ac{ba} graph learn to form a highly efficient influence stream starting from the most central node and passing through its neighbours before reaching more distant neighbours. As the ground-truth value is not important to such agents, this allows the population to quickly coordinate and reach a consensus. Both examples demonstrate non-trivial emergent information sharing structures that would be challenging to know a priori when hand-crafting local agent behaviours as done in most previous literature.

We explicitly demonstrate the benefits of this by comparing our model to some simple hand-crafted baselines. First, we compare the truth-seeking performance of a MARL population ($\alpha=1$) against two myopic Bayesian-inspired heuristics. In both baselines, agents maintain scalar beliefs about the ground truth and select actions that maximise their immediate expected reward.

Each agent $i$ receives a fixed private signal $\obs^i_{(1)}\sim\mathcal{N}(\gtruth,\std^2)$ and begins with a uniform prior, $P(\gtruth=1)=P(\gtruth=0)=0.5$. We represent the agent's belief using its log odds,
\begin{equation}
\ell^i_t=\log\frac{\hat p^i_t}{1-\hat p^i_t},
\label{eq:log_odds}
\end{equation}
where $\hat p^i_t$ denotes its estimated probability that $\gtruth=1$.

Given the uniform prior, Bayes' rule gives the initial log odds,
\begin{equation}
\ell^i_0 =\log \frac{P(\gtruth=1\mid\obs^i_{(1)})}
{P(\gtruth=0\mid\obs^i_{(1)})}=\log
\frac{P(\obs^i_{(1)}\mid\gtruth=1)}
{P(\obs^i_{(1)}\mid\gtruth=0)}=\frac{(\obs^i_{(1)})^2-(\obs^i_{(1)}-1)^2}{2\std^2}=\frac{2\obs^i_{(1)}-1}{2\std^2}.
\end{equation}

At each timestep, we calculate each agent's estimated truth probability from their log odds by rearranging \Cref{eq:log_odds}:
$$
\hat p^i_t=\frac{1}{1+e^{-\ell^i_t}}.
$$

Agents then select an action that maximises immediate expected reward. Given rewards of $+0.1$ for a correct guess, $-0.2$ for an incorrect guess, and $0$ for waiting, the resulting action-selection policy is

\begin{equation}
a^i_t =
\begin{cases}
0, & \hat{p}^i_t \leq \frac{1}{3},\\
1, & \hat{p}^i_t \geq \frac{2}{3},\\
\nullact, & \frac{1}{3}<\hat{p}^i_t<\frac{2}{3}.
\end{cases}
\end{equation}

To incorporate information from neighbouring agents, we treat each observed action as evidence about the ground truth. The evidence associated with an action $a$ is given by its log-likelihood ratio,

\begin{equation}
w_0(a)=\log\frac{P(a_0=a\mid\gtruth=1)}
{P(a_0=a\mid\gtruth=0)},
\end{equation}

where $a_0$ denotes the action of an agent following the initial private-signal-only policy. These action likelihoods are held fixed throughout subsequent interactions. We consider two update rules to incorporate this evidence:

\begin{equation}
\ell^i_{t+1} =
\begin{cases}
\displaystyle \ell^i_0+\sum_{j\in N^\mathrm{out}_\graph(i)}w_0(a^j_t),
& \text{Non-cumulative},\\[8pt]
\displaystyle \ell^i_t+\sum_{j\in N^\mathrm{out}_\graph(i)}w_0(a^j_t),
& \text{Cumulative}.
\end{cases}
\end{equation}

The non-cumulative variant combines the fixed private evidence with the latest neighbour actions, whereas the cumulative variant retains evidence accumulated over previous rounds. Both baselines are Bayesian-inspired heuristics rather than Bayesian inference, as their fixed action likelihoods do not account for changes in neighbours' behaviour or dependencies arising from repeated interactions.

\cref{tab:marl_bayesian_comparison} shows that the MARL population achieves higher mean final accuracy ($0.98$) than both the non-cumulative ($0.89$) and cumulative ($0.87$) baselines. The MARL population also exhibits substantially fewer polarised episodes ($0.01$ compared with $0.37$ and $0.31$, respectively), suggesting that the learned MARL policies are more effective at avoiding persistent disagreement between clusters.

\begin{table}[h]
\centering
\caption{Comparison of MARL with $\alpha = 1$ and Bayesian-based baselines on a 100-agent \ac{sbm} graph evaluated for 200 episodes.}
\label{tab:marl_bayesian_comparison}
\begin{tabular}{lcc}
\toprule
\textbf{Model} &
\makecell{\textbf{Final Population} \\ \textbf{Accuracy}} &
\makecell{\textbf{Fraction of} \\ \textbf{Polarised Episodes}} \\
\midrule
MARL &
\textbf{\boldmath$0.98 \pm 0.01$} &
\textbf{\boldmath$0.01 \pm 0.01$} \\
Non-cumulative Bayes-like &
$0.89 \pm 0.01$ &
$0.37 \pm 0.03$ \\
Cumulative Bayes-like &
$0.87 \pm 0.01$ &
$0.31 \pm 0.03$ \\
\bottomrule
\end{tabular}
\end{table}

We further compare conformity-driven MARL agents ($\alpha=0$)
against three local opinion-dynamics baselines inspired by
DeGroot averaging~\citep{degroot1974reaching}: (i) binary-action
DeGroot, (ii) binary-action DeGroot with a null (``wait'') action,
and (iii) local majority.

Both DeGroot variants maintain a continuous opinion state
$h^i_t\in[0,1]$. Let
$D^i_t=\{j\in N^\mathrm{out}_\graph(i)\mid a^j_t\in\{0,1\}\}$
denote agent $i$'s declaring neighbours. The opinion update is
\begin{equation}
h^i_{t+1}
=
\frac{h^i_t+\sum_{j\in D^i_t}a^j_t}
     {1+|D^i_t|}.
\end{equation}
Thus, the agent gives equal weight to its previous opinion
and each neighbour's non-null declaration. In binary-action
DeGroot, $D^i_t=N^\mathrm{out}_\graph(i)$ and
\begin{equation}
a^i_{t+1}
=
\begin{cases}
0, & h^i_{t+1}<\frac{1}{2},\\
1, & h^i_{t+1}\geq\frac{1}{2}.
\end{cases}
\end{equation}
The variant with waiting instead uses
\begin{equation}
a^i_{t+1}
=
\begin{cases}
0, & h^i_{t+1}\leq\frac{1}{3},\\
1, & h^i_{t+1}\geq\frac{2}{3} ,\\
\nullact, & \frac{1}{3}<h^i_{t+1}<\frac{2}{3}.
\end{cases}
\end{equation}
These thresholds are chosen to be consistent with the MARL conformity rewards: $+0.1$ for agreement, $-0.2$ for disagreement, and $0$ for waiting.

For the local majority update, we define
$n^i_{k,t}=\sum_{j\in N^\mathrm{out}_\graph(i)}\mathbf{1}[a^j_t=k]$, such that the update is
\begin{equation}
a^i_{t+1}
=
\begin{cases}
0, & n^i_{0,t}>n^i_{1,t},\\
1, & n^i_{1,t}>n^i_{0,t},\\
a^i_t, & n^i_{0,t}=n^i_{1,t}.
\end{cases}
\end{equation}

The baselines start from independent initial actions
$a^i_0\sim\mathrm{Bernoulli}(0.5)$, with $h^i_0=a^i_0$
for the DeGroot variants. Waiting becomes available after
the initial declaration.

\cref{tab:marl_degroot_comparison} shows that the convergence speed to consensus is fastest in the MARL population, due to the efficient flow of information from the most central agent to its neighbours and subsequently to more peripheral agents.

\begin{table}[h]
\centering
\caption{Comparison of MARL with $\alpha = 0$ and DeGroot-style baselines on a 100-agent \ac{ba} graph evaluated for 200 episodes.}
\label{tab:marl_degroot_comparison}
\begin{tabular}{lcc}
\toprule
\textbf{Model} & \textbf{Cumulative Reward} &
\makecell{\textbf{First Step Reaching} \\ \textbf{Mean Reward of 0.09}} \\
\midrule
MARL ($\alpha = 0$) & $\mathbf{1.74 \pm 0.01}$ & \textbf{3} \\
Binary DeGroot & $1.28 \pm 0.04$ & Never \\
DeGroot + wait & $1.43 \pm 0.02$ & 10 \\
Local majority & $1.16 \pm 0.05$ & Never \\
\bottomrule
\end{tabular}
\end{table}
\section{Effects of Network Size and Topology} \label[appendix]{app:size_and_topology}

To explore how network size and topology influence resulting dynamics, we compare two network sizes (37 and 1,000 nodes) across two topologies (Bluesky and Hadza). For our comparisons, we construct a 37-node Bluesky network using the same approach described in \cref{app:bluesky_ml}, but with a single source user. We also construct a synthetic 1000-node Hadza-like network designed to preserve the properties of its 37-node counterpart. To do so, we replicate the original degree sequence 27 times, obtaining degrees for 999 nodes, and sample the degree of the final node from the empirical degree distribution. We then randomly generate a graph that preserves this degree sequence, yielding 11\,329 undirected edges. Similarly, we replicate the original set of 419 interaction probabilities 27 times and sample an additional 16 values from the same distribution. These 11\,329 probabilities are then randomly assigned to the edges of the synthetic network. Setting $r=5$ yields an average of 1.19 edges per agent per time step, matching the signal density of the 37-node Hadza network when all agents receive a private signal ($p_\mathrm{signal} = 1$). However, for the 37-agent Bluesky network we instead control for the total number of signals when compared to the 37-agent $r=5$ Hadza network, as this is a more natural way to compare two graphs of the same size.

\begin{figure}[h]
    \centering
    \includegraphics[width=\textwidth]{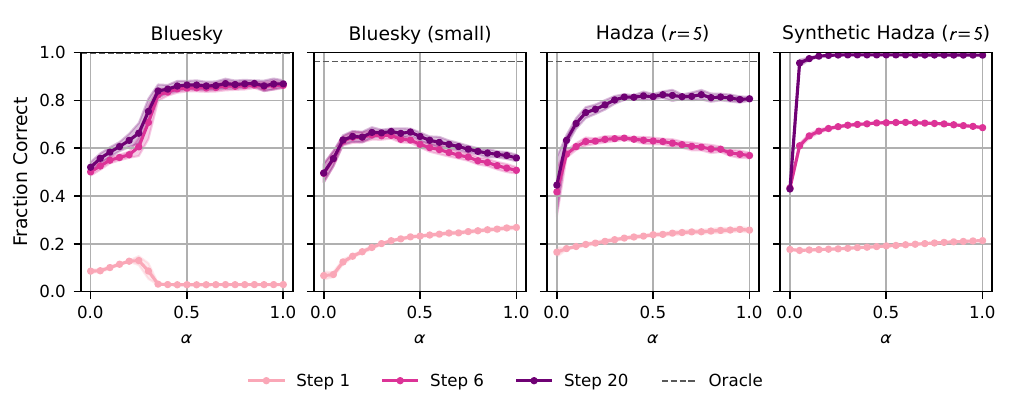}
    \caption{Effect of network size and topology on the fraction of agents whose output matches the ground truth at various time steps for all reward weightings $\weight$ on four distinct graphs. The black dashed line is an oracle that acts Bayes-optimally in isolation given $\lfloor np_\mathrm{signal}\rfloor$ independently sampled private signals. Dynamic Hadza-like networks show higher accuracy than social media networks, and larger networks generally outperform smaller ones.}
    \label{fig:topology_size}
\end{figure}

Given the differences in control variables, comparing the four graphs in \cref{fig:topology_size} requires some care. The most straightforward comparison is between the two 37-node networks. We clearly see that the Hadza topology achieves more correct guesses than the Bluesky topology across all values of $\weight$. This is likely due to the random rewiring effectively allowing agents to gain information from more primary sources throughout an episode, reducing the risk for influential central agents to propagate incorrect information.

Comparing the two Hadza graphs is also reasonably simple, since both have the same signal density, but also share the property that every agent in the population observes a private signal. It is perhaps not surprising to find that a population of 1000 agents outperforms a population of 37 agents, given that the larger population has a much stronger collective signal of the ground truth.

We find a similar result for the two Bluesky networks for $\weight > 0.3$, which again can be explained by the larger population observing approximately four times more private signals than the smaller population. However, we find that for a highly conforming population ($\weight = 0.1$), the smaller population marginally outperforms its larger counterpart. This could be since every agent in the smaller population sees a private signal, giving them a reason to use this information to coordinate on. By contrast, in the larger population, only 14\% of agents get a private signal, meaning that the rest of the population has no signal on which to coordinate for the first time step.

Comparing the two larger networks is less straightforward, as the collective signal is substantially stronger in the synthetic Hadza network, resulting in near-perfect final accuracy. However, the Bluesky network achieves higher accuracy at earlier time steps, indicating faster initial convergence. Crucially, as discussed in \cref{sec:accuracy}, the 37-node Hadza network performs comparably to the 1000-node Bluesky network in the truth-seeking case and outperforms it in the conformity-driven case, despite having four times fewer private signals overall. These findings suggest that the Hadza topology offers advantages for collective information propagation over both social media topologies.

\begin{figure}[h]
    \centering
    \includegraphics[width=0.85\textwidth]{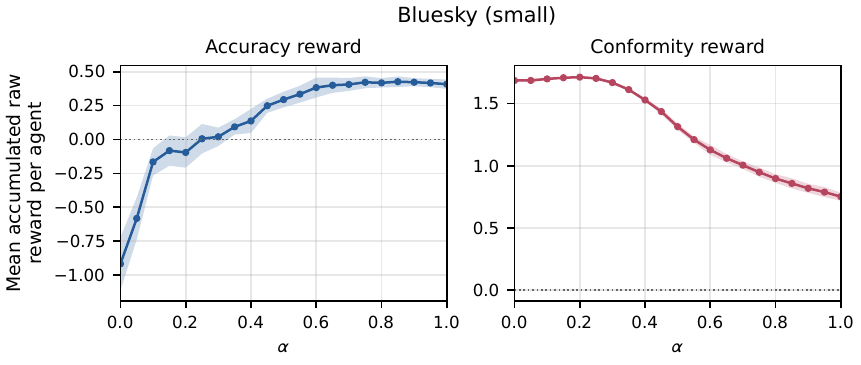}
    \caption{Raw rewards before weighting for all $\weight$ for the 37-agent Bluesky network. High $\weight$ optimises for accuracy while low $\weight$ optimises for conformity.}
    \label{fig:raw_rewards_bluesky_small}
\end{figure}

\newpage

An interesting phenomenon emerges in smaller networks, whereby the fraction of correct guesses can decrease with increasing truth-seeking incentives, particularly at intermediate time steps. While this may initially seem counterintuitive, it can be explained by the asymmetric reward function, which penalises incorrect guesses more heavily than it rewards correct ones (-0.2 compared to +0.1). Consequently, agents are incentivised to wait whenever their posterior belief that the ground truth takes a particular value lies between 1/3 and 2/3. As a result, stronger truth-seeking incentives can lead to more agents choosing to wait rather than risk an incorrect guess, thereby reducing the overall fraction of correct guesses. \cref{fig:raw_rewards_bluesky_small} confirms that this is not an optimisation failure, but rather an emergent property of the dynamics,  as populations with high $\weight$ are still maximising their raw individual accuracy rewards.

\section{Model Robustness} \label[appendix]{app:model_robustness}

\begin{figure}[h]
    \centering
    \includegraphics[width=\textwidth]{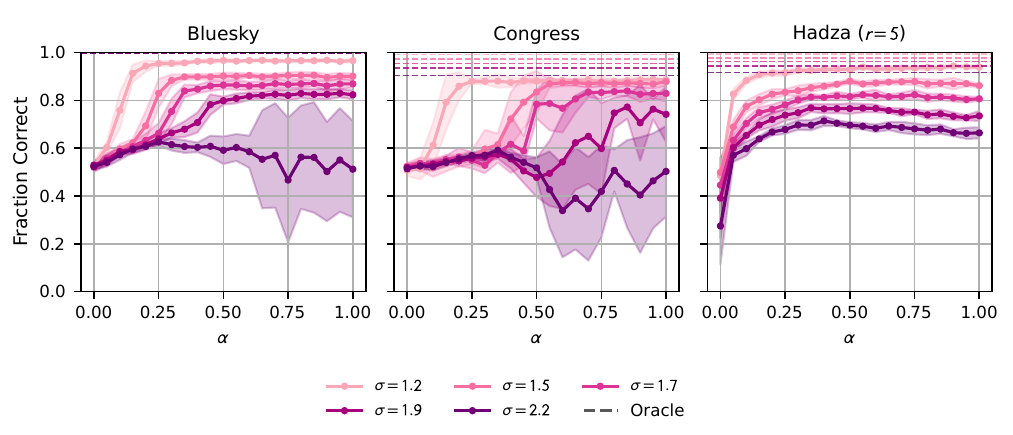}
    \caption{Effect of changing the private signal uncertainty $\std$ on the fraction of agents whose final output matches the ground truth for all reward weightings $\weight$ on three distinct graphs. The dashed lines are oracles that act Bayes-optimally in isolation given $\lfloor np_\mathrm{signal}\rfloor$ independently sampled private signals. Larger social media networks suffer more from noisier signals than smaller networks.}
    \label{fig:sigma_sweep}
\end{figure}

To assess the robustness of our setup, we examine how learning depends on the uncertainty in the ground truth. When uncertainty is low, agents have little incentive to communicate, whereas high uncertainty provides too weak a signal for effective learning. As shown in \cref{fig:sigma_sweep}, increasing $\std$ reduces population accuracy and can substantially destabilise learning in larger social media networks. We therefore choose $\std = 1.7$, which allows us to study all networks under reasonably stable learning conditions while representing a human population coordinating on a task that is challenging but not impossible to solve.

\begin{figure}
    \centering
    \includegraphics[width=\textwidth]{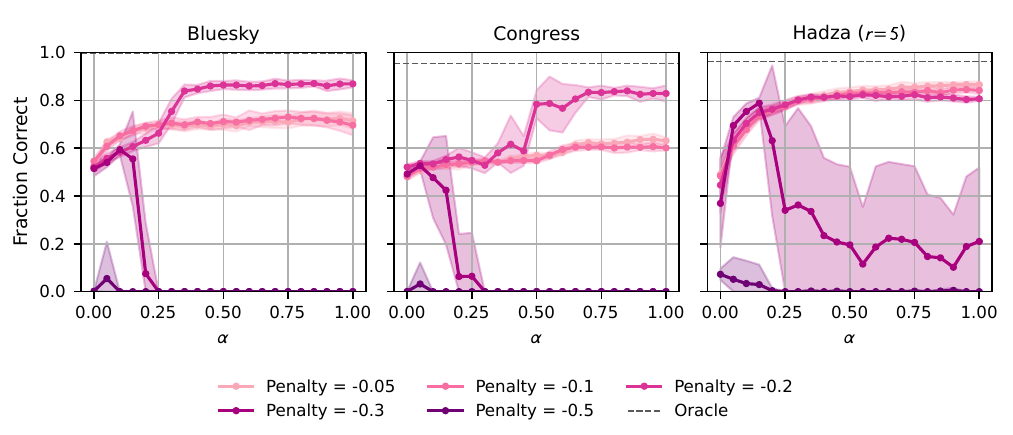}
    \caption{Effect of changing the reward symmetry on the fraction of agents whose final output matches the ground truth for all reward weightings $\weight$ on three distinct graphs. Positive and neutral rewards are kept constant at 0.1 and 0. The dashed lines are oracles that act Bayes-optimally in isolation given $\lfloor np_\mathrm{signal}\rfloor$ independently sampled private signals. Large penalties result in agents indefinitely using the null action.}
    \label{fig:reward_sweep}
\end{figure}

We additionally explore how the asymmetry of our reward function affects the resultant learning dynamics. \cref{fig:reward_sweep} shows results for different penalty values, while keeping positive rewards at 0.1 and neutral rewards at 0. We refer to a reward function that has a positive (negative) expected reward over random actions as positively (negatively) shifted, as opposed to symmetric when this expectation is zero. We find that positively shifted and symmetric rewards (penalties of -0.05 and -0.1, respectively) result in better accuracy than negatively shifted rewards on the small network, whereas the opposite is true for the two larger networks. This can be explained as follows: in the larger social media networks, only a small fraction of agents receive a private signal. If the reward is positively shifted, every agent has an incentive to guess something, even if they do not observe a private signal. This creates extra noise in the network, making it more difficult for genuine signals to propagate through the population. By contrast, since smaller networks can collectively suffer under a negatively shifted reward when too many individual agents use the wait action, particularly for truth-seeking populations ($\weight = 1$), changing this to a positively shifted reward results in all agents outputting guesses, creating a stronger collective signal of the ground-truth value.

For large negatively shifted rewards, all populations resort to using the null action indefinitely, as making a  wrong guess or disagreeing with a neighbour is highly costly.

\section{Ablations} \label[appendix]{app:ablations}

\begin{figure}[htbp]
    \centering
    \includegraphics[width=\textwidth]{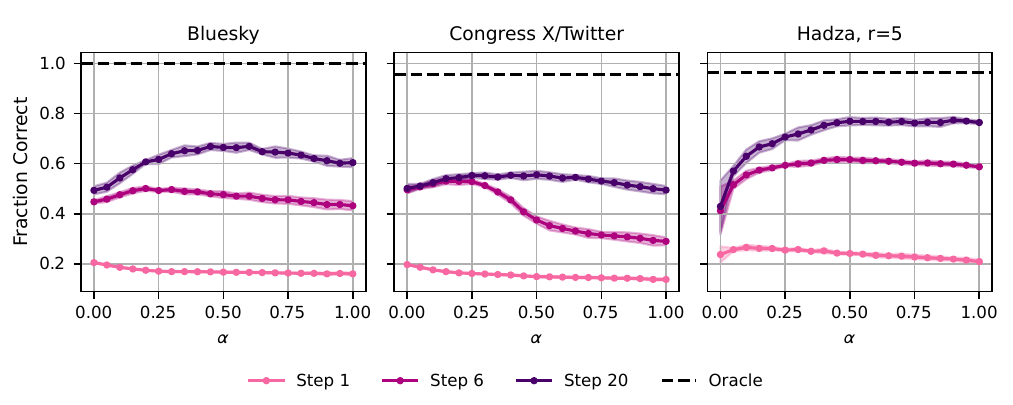}
    \caption{Effect of ablating GRU on the fraction of agents whose output matches the ground truth at various time steps for all reward weightings $\weight$ on three distinct graphs. The black dashed line is an oracle that acts Bayes-optimally in isolation given $\lfloor np_\mathrm{signal}\rfloor$ independently sampled private signals. Ablating the GRU decreases the performance of static graphs significantly.}
    \label{fig:ablation_rnn}
\end{figure}

\begin{figure}[htbp]
    \centering
    \includegraphics[width=\textwidth]{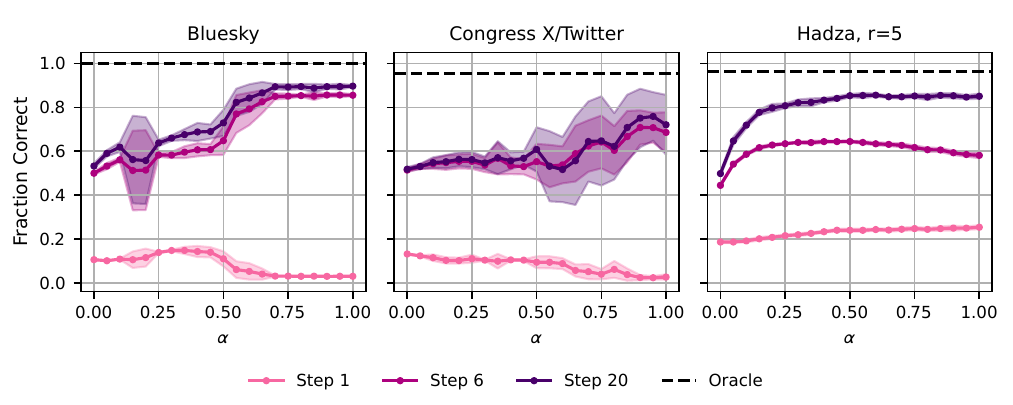}
    \caption{Effect of ablating agent IDs on the fraction of agents whose output matches the ground truth at various time steps for all reward weightings $\weight$ on three distinct graphs. The black dashed line is an oracle that acts Bayes-optimally in isolation given $\lfloor np_\mathrm{signal}\rfloor$ independently sampled private signals. Ablating agent IDs decreases the performance of static graphs significantly.}
    \label{fig:ablation_ids}
\end{figure}

\Cref{fig:ablation_rnn,fig:ablation_ids} show the effects of ablating the GRU component and agent IDs, respectively, on the accuracy of the population. We see that the performance is significantly reduced on static graphs such as the Bluesky and Congress X/Twitter graphs, but that the effects are less pronounced for the dynamic Hadza graph. This highlights the importance of the time axis when repeatedly communicating with the same neighbours, and the benefits of knowing which neighbours are sending which signals. Agents on the Hadza network rely less on these aspects due to the stochastic nature of interactions with the rest of the population.

\begin{figure}[htbp]
    \centering
    \includegraphics[width=\textwidth]{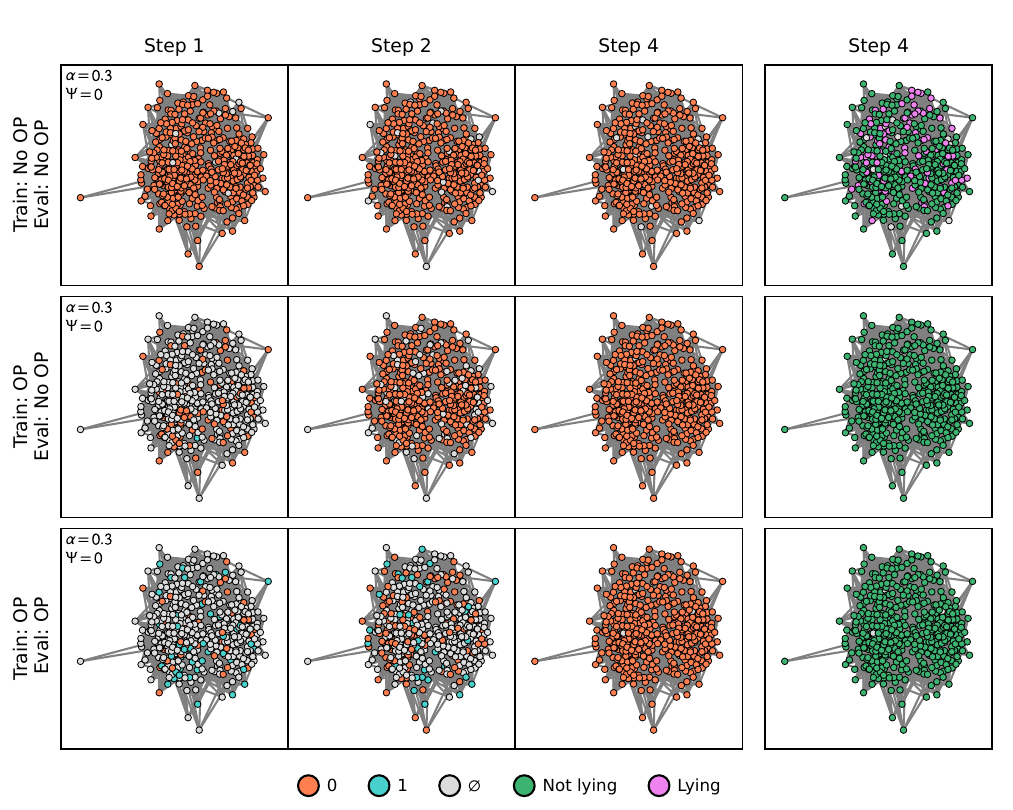}
    \caption{Ablation examples for the Congress network. \textbf{Top row:} Population trained and evaluated without \ac{op}. \textbf{Middle row:} Population trained with \ac{op} and evaluated without \ac{op}. \textbf{Bottom row:} Population trained and evaluated with \ac{op}. Both training and evaluating with \ac{op} are required to prevent modelling with unrealistic conventions.}
    \label{fig:ablation}
\end{figure}

\FloatBarrier

\Cref{fig:ablation} demonstrates why we use \ac{op} for both training and evaluation in our model. In this example, training and evaluating without \ac{op} leads to the fairly conforming population forming a convention to always output 0 regardless of their private signal values. This only aligns with the ground truth half of the time. This maximises the conformity component of the reward but is undesirable for modelling information diffusion. If we train with \ac{op} and evaluate without \ac{op}, as is usual for evaluating policy performance in the literature of \ac{zsc}, we find that agents can learn to use the symmetry flipping of \ac{op} to enforce stochasticity in their initial outputs if they do not use the null action, despite the underlying policy for initial non-null outputs actually collapsing to 0 (remember this output is then flipped by \ac{op} in training). Thus, there is an accidental convention to always converge to 0, which again has an accuracy of 50\%.
Finally, training and evaluating with \ac{op} gives rise to desired behaviours of information diffusion.

\section{Testing for Nash Equilibrium} \label[appendix]{app:nash}

\begin{figure}[h]
    \centering
    \includegraphics[width=\textwidth]{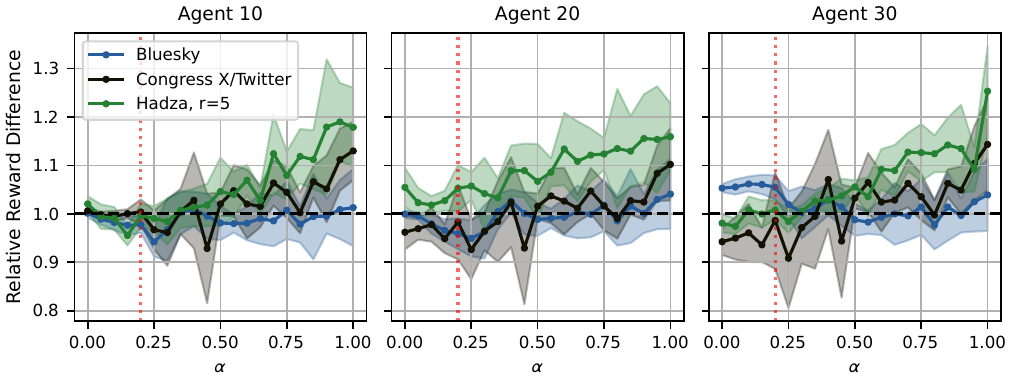}
    \caption{Relative reward difference for three randomly selected agents retrained in separate tests to exploit unilateral deviation against a previously trained population. Reward gains of up to 20\% for large $\weight$ indicate that the model does not fully converge to \ac{ne}.}
    \label{fig:local_nash_3}
\end{figure}

To test for \ac{ne}, we keep a frozen copy of parameters and train a single randomly selected agent on a newly initialised set of parameters. If the agent cannot improve its expected return, this indicates evidence of a \ac{ne}. To fully test for \ac{ne}, we would need to repeat this for every agent, which is computationally expensive for large populations. We thus restrict this to just three randomly selected agents, which proves sufficient for our analysis. \Cref{fig:local_nash_3} shows the average relative reward difference for three randomly selected agents before and after retraining in separate tests. We find that the average can increase by up to 20\% for larger $\weight$. This demonstrates that agents do not fully converge to a \ac{ne}, though they do converge to stable policies (see \Cref{app:hyperparameters}). This is somewhat unsurprising given that even a more constrained form of multi-agent learning known as regret minimisation is only guaranteed to reach a \ac{cce}~\citep{blum2008regret}. It is also debated whether human behaviour can be reliably predicted by \ac{ne}~\citep{nagel1995unraveling,mckelvey1995quantal}, while reinforcement learning dynamics have been shown to provide more accurate predictions~\citep{erev1998predicting}. Thus, these results do not invalidate our model.

\section{Interpreting Learned Policies} \label[appendix]{app:interpreting_learned_policies}

\begin{figure}[htbp]
    \centering
    \includegraphics[width=\textwidth]{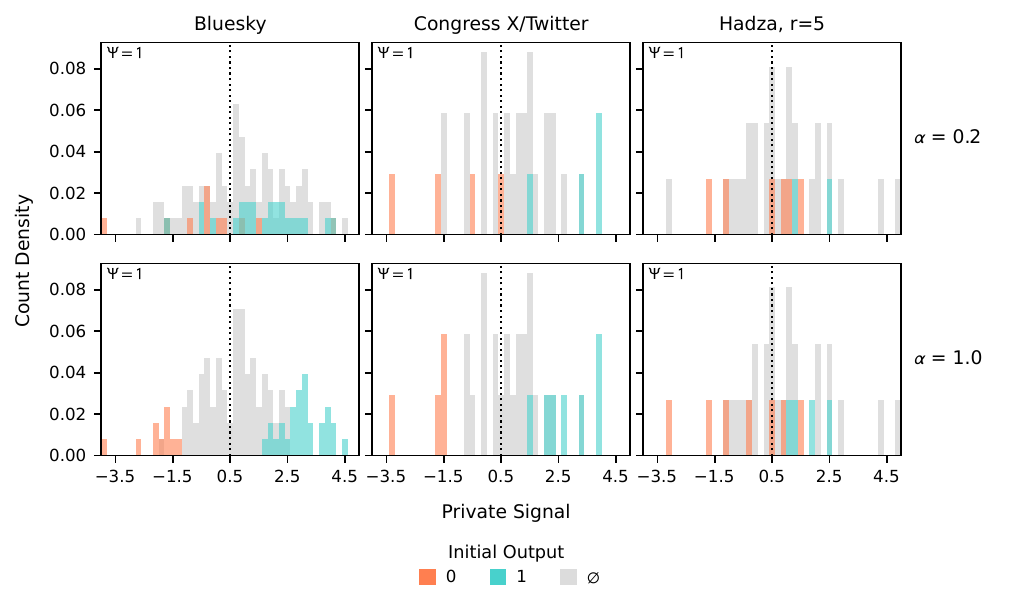}
    \caption{Histograms showing the distribution of initial outputs for agents receiving a private signal in a selected episode on three distinct graphs. \textbf{Top row:} $\weight = 0.2$. \textbf{Bottom row:} $\weight = 1.0$. Truth-seeking agents ($\weight$ = 1.0) in the social media graphs develop a strategy to withhold their output on the first turn when  unsure of their private signal.}
    \label{fig:initial_guess_hist}
\end{figure}

\begin{figure}[htbp]
    \centering
    \includegraphics[width=\textwidth]{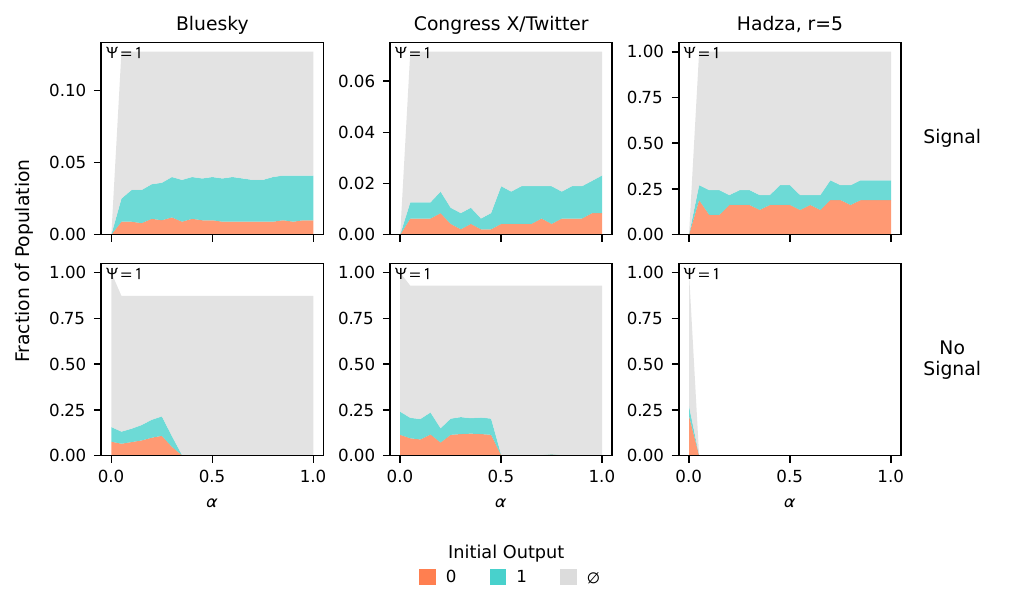}
    \caption{Area charts showing fractional distribution of initial outputs for all values of $\weight$. \textbf{Top row:} Agents receiving a private signal. \textbf{Bottom row:} Agents not receiving a private signal. Conforming populations lead to an increase in outputs from agents not receiving private signals.}
    \label{fig:initial_guess_area_chart}
\end{figure}

Since the policies learned by agents are complex, it is difficult to gain a detailed understanding of agents' actions at every time step. However, we can still interpret some of the learned policies by analysing the actions at the initial time step. \Cref{fig:initial_guess_hist} shows how the initial output frequencies differ between a highly conforming population ($\alpha = 0.2$) and a fully truth-seeking population ($\alpha = 1$) for agents that receive a private signal. We see that in the social media graphs, truth-seeking agents withhold their guesses on the first time step when uncertain of their beliefs (i.e. when their private signals are near 0.5), while conforming agents are not so strict. Interestingly, this region of withholding does not emerge in the Hadza network, which is likely caused by the small population size and limited communication increasing the value of each agent's individual guess. This does not mean that there is no withholding happening in the Hadza network, which mainly occurs during intermediate steps (\Cref{fig:accuracy}).

To understand the behaviour of agents without a private signal, we can analyse the distribution of initial outputs as a function of $\weight$, as seen in \Cref{fig:initial_guess_area_chart}. For the social media networks, which are the only graphs that have a fraction of agents without private signals, we see that the number of non-null guesses increases for $\weight < 0.5$, indicating that agents care more about fast conformity than waiting for rare signals from agents that receive private signals of the truth, $\gtruth$. For agents that do receive a private signal, we see that the larger fraction of non-null initial outputs consistently aligns with $\Psi$ for all $\weight$, except for the Hadza network, which by chance had more incorrect guesses. This is expected since, on average, more agents will receive a signal that aligns their posterior beliefs with the true value $\Psi$.

\section{Extracting Attention Weights} \label[appendix]{app:extracting_attention_weights}

\begin{figure}[h]
    \centering
    \includegraphics[width=0.9\textwidth]{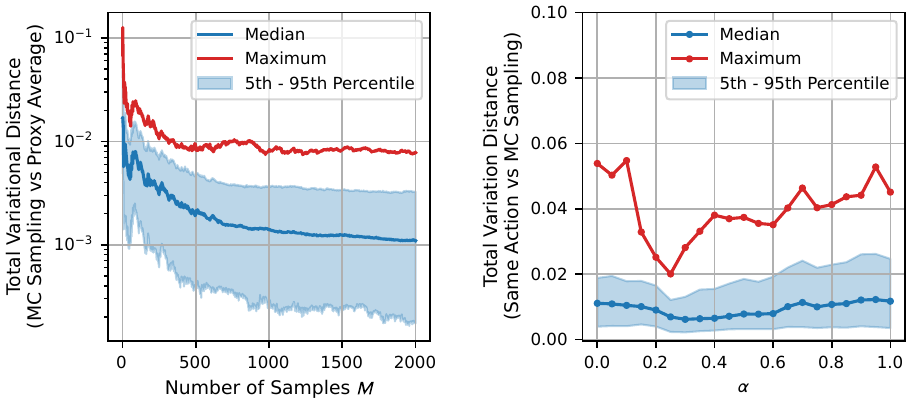}
    \caption{\textbf{Left:} TV distance between attention weights estimated using increasing numbers of MC samples and a reference estimate computed using $10^4$ MC samples. \textbf{Right:} TV distance between attention weights obtained via MC sampling and the same-action method, evaluated across different values of $\weight$. Both experiments use the 1000-agent Bluesky network.}
    \label{fig:attention_extraction}
\end{figure}

Since $\act_{t-1}$ is contained in $\statet$, $\obsfunc$ depends only on $\state$, and we can approximate the expected attention via Monte Carlo (MC) sampling of episode trajectories,
\begin{equation}
\bar{w}_{ij} = \mathbb{E}_{\state \sim \rho^\policy}\left[\mathbb{E}_{\obs^i \sim \obsfunc^i(\cdot|\state)}\left[w_{ij}(\obs^i)\right]\right] \approx \frac{1}{M}\sum_{m=1}^{M} w_{ij}(\obs^{i,(m)}).
\end{equation}
We need to be confident that we are taking enough samples for a good estimate, and thus analyse convergence of the total variation (TV) distance over attention distributions for each agent, given by
\begin{equation}
    \mathrm{TV}_i = \frac{1}{2} \sum_{j \in N^\mathrm{out}_\graph[i]} \left| \bar{w}_{ij} - \bar{w}^*_{ij} \right|,
\end{equation}
where $\bar{w}_{ij}^*$ are proxy true weights taken with $M=10^4$. \Cref{fig:attention_extraction} (left) demonstrates that the TV distance converges to within 1\% for all agents when $M>500$. Our evaluations are taken with 4000 steps (200 episodes), hence we can be confident that our calculated attention weights are an accurate representation of $\bar{w}_{ij}^*$.

MC sampling is not the only way in which we could extract attention weights. For example, we could also apply a simple heuristic by setting all previous actions equal and observing the effects of the agent IDs when all agents are in agreement. \Cref{fig:attention_extraction} (right) shows the TV distance between weights calculated via MC sampling vs the simple heuristic of equal actions. We see that the median TV distance lies around 1\% with the worst case around 4\%. The choice of method for extracting attention weights does not significantly impact any of our results.

\section{Centrality Measures} \label[appendix]{app:centrality_measures}

Two common attention-based centrality measures used in the context of graphs include the sum over received attentions and the average received attention, given by
\begin{equation}
C_\mathrm\Sigma(i) = \sum_{j\in N^\mathrm{in}_\graph[i]}w_{ji}
\end{equation}
and
\begin{equation}
C_\mu(i) = \frac{1}{|N^\mathrm{in}_\graph[i]|}\sum_{j\in N^\mathrm{in}_\graph[i]}w_{ji},
\end{equation}
respectively. Both measures have a natural bias to giving high degree nodes large centrality. For example, imagine we have a star graph with 100 nodes $m_i$ for $1\leq i \leq 100$ where a single central node $m_1$ has 99 edges, one connecting directly to every other node. If all edges have uniform attention weights, all peripheral nodes give a weighting of 0.5 to themselves and to the central node, while the central node gives a weighting of 0.01 to itself and all peripheral nodes. For the central node this gives centrality scores of $C_\Sigma(1) = 49.51$ and $C_\mu(1) = 0.4951$, while for each peripheral node $1<j\leq100$ we get scores of $C_\Sigma(j) = 0.51$ and $C_\mu(j) = 0.255$.

While such a bias might be desirable in determining which nodes might have the greatest influence over the network, we are more interested in the \textit{perceived} importance at the individual agent level. If all agents uniformly weight their closed neighbourhood, then all agents are perceived as equally important and so should receive the same perceived importance score. We do this by multiplying a given score by the corresponding agent's out-degree, so that a score of 2 means that this agent perceives the selected neighbour to be twice as important relative to uniform attentions. We can then safely average over the received importance scores to arrive at
\begin{equation}
C_\mathrm{PI}(i) = \frac{1}{|N^\mathrm{in}_\graph[i]|}\sum_{j\in N^\mathrm{in}_\graph[i]} |N^\mathrm{out}_\graph[j]|w_{ji},
\end{equation}
which is \Cref{eq:pi} in the main text. If we repeat the example of the star graph with uniform weights we see that we get desired scores of $C_\mathrm{PI}(i) = 1$ for all $1\leq i \leq 100$.
\section{Edge Weight Baselines} \label[appendix]{app:edge_weight_baselines}

We use five heuristic baselines for predicting edge weights used by \citet{chandrasekhar2015testing} in the context of DeGroot learning~\citep{degroot1974reaching} and \citet{liben2003link} in the context of link prediction. These include uniform weighting, degree weighting, eigenvector weighting, Jaccard and Adamic-Adar. We use $A'=A + I$, where $I$ is the identity matrix, to denote the adjacency matrix with self-loops on all nodes.

\textbf{Uniform weighting.} Here an agent places all weighting uniformly among its neighbours and itself. The weight matrix is given by
\begin{equation}
w^U_{ij} = \frac{A'_{ij}}{d^\mathrm{out}_i},
\end{equation}
where $d^\mathrm{out}_i$ is agent $i$'s out-degree (including itself).

\textbf{Degree weighting.} Here each agent places a weighting proportional to their neighbours' (and self) popularity. This is given by
\begin{equation}
w^D_{ij} = \frac{d^\mathrm{in}_j}{\sum_{k \in N^\mathrm{out}_\graph[i]} d^\mathrm{in}_k},
\end{equation}
where $d^\mathrm{in}_i$ is agent $i$'s in-degree (including itself).

\textbf{Eigenvector weighting.} Here the weighting is determined by the eigenvector centrality, determined by the left eigenvector of the adjacency matrix with the largest eigenvalue. More formally this is given by
\begin{equation}
w^E_{ij} = \frac{\xi_j}{\sum_{k \in N^\mathrm{out}_\graph[i]} \xi_k},
\end{equation}
where $\xi_i$ satisfies $\sum_i\xi_iA'_{ij} = \lambda^*\xi_j$ and $\lambda^*$ is the largest eigenvalue.

\textbf{Jaccard.} Here the weighting between two agents is determined by the Jaccard coefficient,
\begin{equation}
w^J_{ij} = \frac{|N^\mathrm{out}_\graph(i) \cap N^\mathrm{out}_\graph(j)|}{|N^\mathrm{out}_\graph(i) \cup N^\mathrm{out}_\graph(j)|},
\end{equation}
which weights agents with a similar out-neighbourhood more highly.

\textbf{Adamic-Adar.} Here the edges between two agents are weighted similarly to Jaccard, but we divide by the popularity of the joint neighbours instead of the union of the neighbourhoods. This is given by
\begin{equation}
w^{AA}_{ij} = \sum_{k \in N^\mathrm{out}_\graph(i) \cap N^\mathrm{out}_\graph(j)} \frac{1}{\log d_k^\mathrm{in}},
\end{equation}
where in this case the in-degree does not include itself.

We note that the uniform, degree and eigenvector baseline weight matrices are derived using an adjacency matrix with self-loops on all nodes to match how the \ac{marl} model allows agents to observe their own previous actions. However, just as we do for the learned \ac{marl} weights, we remove the diagonal (self weights) and renormalise the rows for fair comparison with the real data, which do not contain self-loops.

\section{Validation Metrics} \label[appendix]{app:validation_metrics}

\begin{figure}[h]
    \centering
    \includegraphics[width=\textwidth]{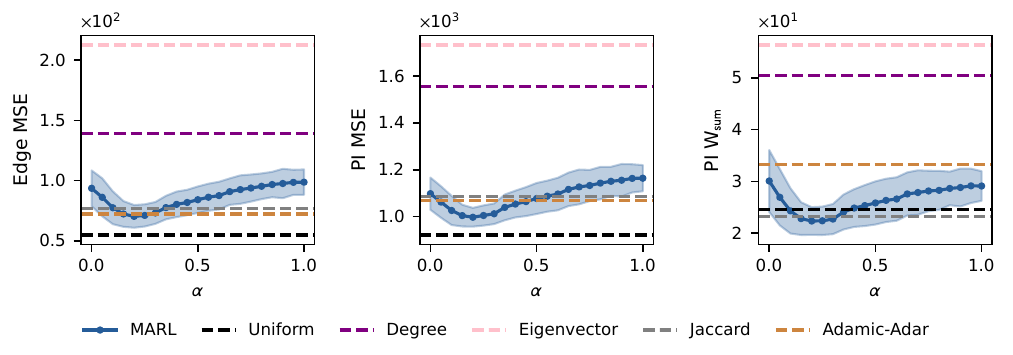}
    \caption{Three validation metrics, from left to right: edge-level \ac{mse}, node-level \acl{pi} \ac{mse}, \ac{pi} Wasserstein summation over node in-degree. Uniform weights score highly on basic statistics (\ac{mse}) but worse in the $W_\mathrm{sum}$ metric as they fail to capture the distribution of node importances.}
    \label{fig:validation_metrics}
\end{figure}

\begin{figure}[h]
    \centering
    \includegraphics[width=\textwidth]{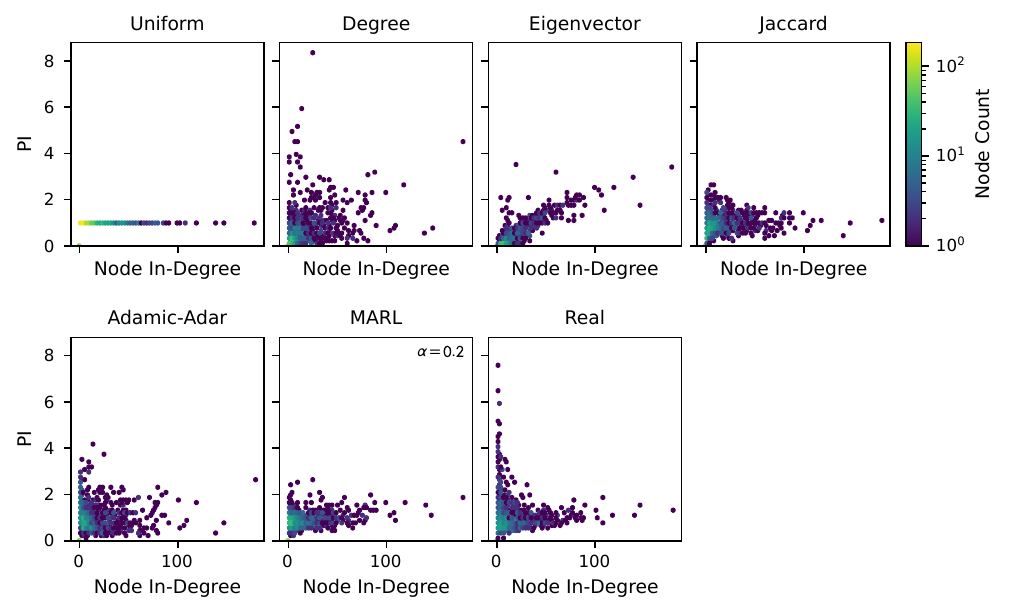}
    \caption{Hexagonal histograms for the Bluesky network showing \ac{pi} distributions as a function of node in-degree. \textbf{Top row:} uniform weighting, degree weighting, eigenvector weighting and Jaccard. \textbf{Bottom row:} Adamic-Adar, \ac{marl} predictions for $\weight = 0.2$ and the real data. The distribution generated by the \ac{marl} simulation is visually more accurate than the baselines.}
    \label{fig:validation_bluesky_all_baselines}
\end{figure}

Similarity metrics such as \ac{mse} and cosine similarity offer a simple way of comparing edge weights between two graphs. However, they fail to correctly capture distributional differences beyond the mean. For example, in the case of the Bluesky dataset, uniform weights score highly on \ac{mse}-based metrics (\Cref{fig:validation_metrics}), since the average \ac{pi} is roughly uniform (\Cref{fig:validation_bluesky_all_baselines}). However, from \Cref{fig:validation_bluesky_all_baselines} we see that the uniform weights do not account for the distribution in \ac{pi} scores observed in the real data. We thus calculate the Wasserstein distance for each node in-degree and sum to obtain a scalar metric, $W_\mathrm{sum}$, that better captures the distribution. We see that this metric results in the Jaccard and \ac{marl} weights scoring better than the uniform weights (\Cref{fig:validation_metrics}), as they better capture the width of the distribution (\Cref{fig:validation_bluesky_all_baselines}).

\FloatBarrier

\section{Validating the Congress X/Twitter Network}\label[appendix]{app:validating_congress}

\begin{figure}[h]
    \centering
    \includegraphics[width=\textwidth]{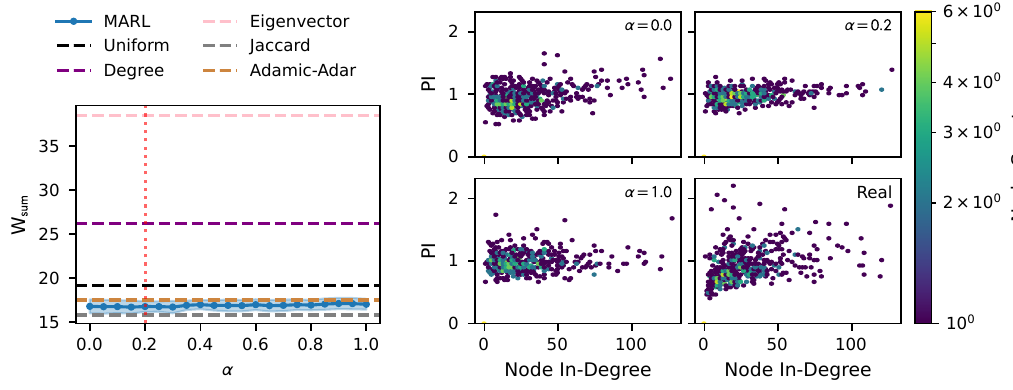}
    \caption{\textbf{Validation of Congress network.} \textbf{Left:} Sum of Wasserstein distances for each node in-degree. \textbf{Right grid:} Hexagonal histograms showing \ac{pi} distributions as a function of node in-degree, comparing \ac{marl} prediction at $\weight =$ 0, 0.2, and 1 with the real data. The \ac{marl} simulations struggle to capture the full shape of the real data.}
    \label{fig:validation_congress}
\end{figure}

\begin{figure}[h!]
    \centering
    \includegraphics[width=\textwidth]{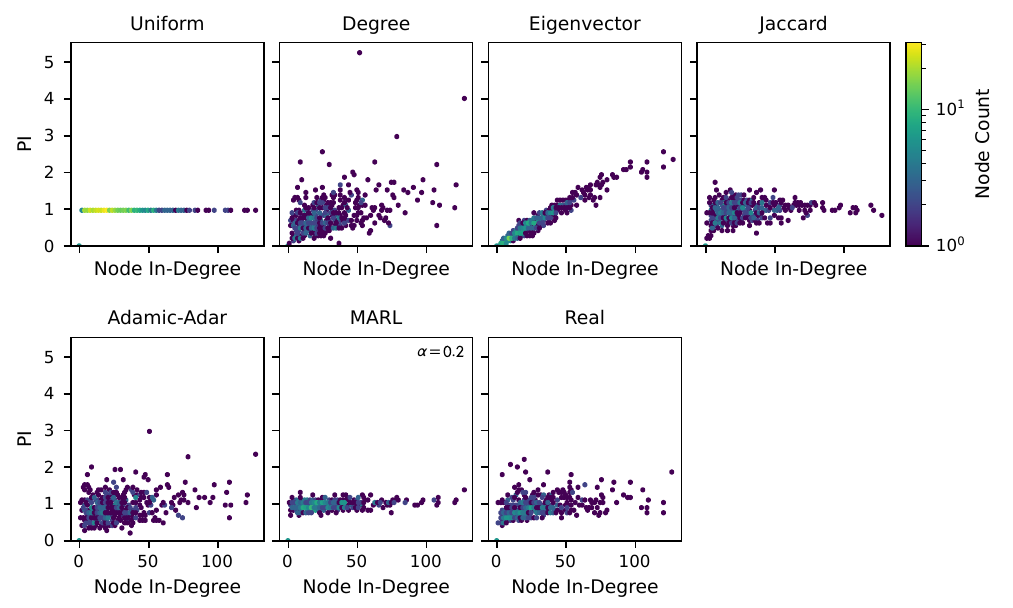}
    \caption{Hexagonal histograms for the Congress X/Twitter network showing \ac{pi} distributions as a function of node in-degree. \textbf{Top row:} uniform weighting, degree weighting, eigenvector weighting and Jaccard. \textbf{Bottom row:} Adamic-Adar, \ac{marl} predictions for $\weight = 0.2$ and the real data. None of the predicted distributions visually fit the real data well.}
    \label{fig:validation_congress_all_baselines}
\end{figure}

While our model seems to predict \ac{pi} structures fairly well for the Bluesky dataset, it cannot predict edge weights for any arbitrary graph, particularly when actors behave with complex incentives. \Cref{fig:validation_congress,fig:validation_congress_all_baselines} show the \ac{pi} fit for the empirical Congress X/Twitter dataset. We see that the learned \ac{marl} weights do not capture the distribution accurately. One reason for this is that real humans do not just act according to conformity and truth-seeking incentives, especially politicians. They tend to have additional motives such as political influence and reputation. Our model is not designed to capture such behaviours. We choose to include this graph in our analyses as a good example of emergent dynamics on a clustered topology.

\section{Notions of Accuracy} \label[appendix]{app:notions_of_accuracy}

\begin{figure}[htbp]
    \centering
    \includegraphics[width=\textwidth]{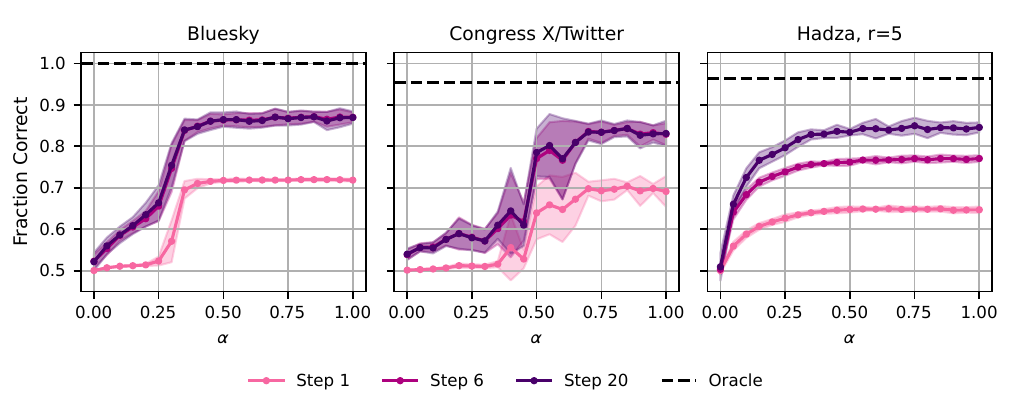}
    \caption{Fraction of agents among the non-null outputting subpopulation whose output matches the ground truth at various time steps for different $\weight$ on three distinct graphs. The black dashed line is an oracle that acts Bayes-optimally in isolation given $\lfloor np_\mathrm{signal}\rfloor$ independently sampled private signals. Under this alternative accuracy definition, finite conformity is no longer advantageous for the Hadza network during intermediate timesteps.}
    \label{fig:accuracy_guesses_only}
\end{figure}

In \Cref{sec:accuracy}, we define population accuracy as the fraction of all agents making a correct guess. However, we can also consider only agents that make a non-null guess (i.e. not $\emptyset$). Under this alternative measure, the Hadza network no longer scores higher for finite conformity during intermediate timesteps (\Cref{fig:accuracy_guesses_only}). This measure might be more appropriate for a vote, where agents that remain silent do not count. In the main text, however, we consider it important that agents both make a guess and that this guess is correct, so the increase in correct guesses under finite conformity in the Hadza network still represents an improvement.

\section{Validating the Belief Head} \label[appendix]{app:belief_head}

\begin{figure}[h]
    \centering
    \includegraphics[width=\linewidth]{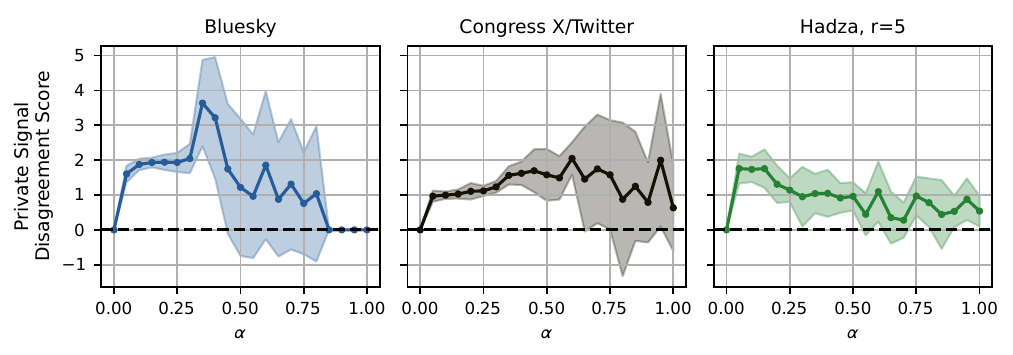}
    \caption{Average \acl{psd} scores for lying agents receiving private signals for different $\weight$ on three distinct graphs. When $\weight > 0.5$, we get very few lying agents (\Cref{sec:dishonesty}), so the belief head is valid without needing to look at \ac{psd} scores. For $0 < \weight < 0.5$, we find \ac{psd} scores to be largely positive, indicating that the belief head correctly opposes the population guess when an agent's private signal induces a strongly opposing posterior. When $\weight = 0$, the \ac{psd} scores drop to zero since agents do not receive private signals for this value of $\weight$.}
    \label{fig:belief_head_validation}
\end{figure}

In order to draw conclusions about dishonest agents, we need to ensure that the belief head has been trained sufficiently. However, since the population accuracy can be quite low for mostly conforming agents ($\weight < 0.5$), the belief loss will be noisy, even if it gives a best possible prediction of the underlying truth state from the agent's hidden state. We verify whether the belief output behaves as expected using a \ac{psd} score,
$$
\mathrm{PSD}(i) = \obs^i_{(1)} - 0.5 - \act_{H-1}^{\mathrm{maj}} (2\obs^i_{(1)} - 1),
$$
where $\act_{H-1}^{\mathrm{maj}} \in \{0,1\}$ is the majority non-null action of the population in the final round of guessing. A positive score means that an agent's private signal disagrees with the final majority guess. For example, if $\act_{H-1}^{\mathrm{maj}} = 0$, all private signals greater than 0.5 give positive disagreement scores, while for $\act_{H-1}^{\mathrm{maj}} = 1$, all private signals less than 0.5 give positive disagreement scores. The strength of these scores is determined by how far the private signals lie away from 0.5.
When $\weight > 0.5$, there are very few lying agents (\Cref{sec:dishonesty}), which naturally validates the behaviour of the belief head, as agents act according to what they believe to be true. When $0 <\weight < 0.5$, we would expect lying agents that conform to the population guess to have a private signal that strongly opposes this, and thus a large positive \ac{psd} score. \Cref{fig:belief_head_validation} shows that such agents do in fact have large positive \ac{psd} scores, thus validating the behaviour of the belief head for this range of $\weight$. Finally, when $\weight = 0$, we see that the \ac{psd} scores drop to zero as expected, since agents do not receive private signals for this value of $\weight$.

\section{More on Private Signal Densities}\label[appendix]{app:signal_densities}

Since the Bluesky and Congress X/Twitter networks have higher edge densities than the Hadza tribe network, agents observe proportionally more information from their neighbours during the initial round of guessing. To control for this, we add sparsity to the private signals by introducing a probability $p_\mathrm{signal}$ of observing a signal, where a ``no signal'' case gives agents an uninformative private signal of 0.5. During training, we start from unity and decay this probability exponentially for increased learning stability. The final probability is calculated by matching the signal density to that of the Hadza network with $r=5$. For example, the Bluesky network has 1000 nodes and 14\,559 directed edges, giving an average of 14.6 neighbouring signals per node, or 15.6 total signals including private signals. Similarly, the Twitter network has 475 nodes and 13\,289 edges, giving an average of 28 neighbouring signals per node, or 29 total signals including private signals. Comparing to the Hadza graph with $r=5$, which has an average of 2.19 total signals per node (37 nodes and an average of 44.1 edges), we can set $p_\mathrm{signal}(\mathrm{Bluesky}) = 0.14$ and $p_\mathrm{signal}(\mathrm{X/Twitter}) = 0.08$ to standardise the total signal density to be equal to that of the Hadza network with $r=5$.

There are different quantities that we could hold constant across graph sizes, such as the variance in the private signal $\std^2$ or the number of agents that receive a signal. However, we choose this quantity as it helps to standardise the strength of neighbouring signals on the first time step of each episode for a single agent.
\section{Hyperparameters and Training Curves} \label[appendix]{app:hyperparameters}

The hyperparameters we use during training are given in \Cref{tab:training_parameters}. We use linear annealing for the learning rate with a final learning rate of zero. For tuning, we choose $\std$ (private signal uncertainty) to provide a sufficiently weak signal that requires agents to share information over multiple time steps, while remaining below the learnability threshold during training (\Cref{app:model_robustness}). We find the entropy coefficient to have a significant effect on the learning for truth-seeking agents, with greater entropy improving learnability during training. This is tuned to the lowest value for which $\weight = 1.0$ populations could learn without collapsing.

For $\weight = 0$, we choose to set $p_\mathrm{signal} = 0$ for all agents to avoid learning additional unwanted conventions that are \ac{op} invariant and only affect this value of $\weight$. For numerical stability, we do this by decaying exponentially to $\varepsilon = 0.01$ after which we decrease linearly to 0. Additionally, we find that $\weight = 0$ frequently gets stuck in suboptimal policies, which we solve by shaping $\weight$ linearly from 0.05 to 0 during early stages of training.

\vspace{6em}

\begin{table}[h]
\centering
\caption{Hyperparameters used for training.}
\label{tab:training_parameters}
\begin{tabular}{lc}
\toprule
\textbf{Hyperparameter} & \textbf{Value} \\
\midrule
$H$ & 20 \\
Number of Parallel Environments & 20 \\
Total Training Timesteps & $10^6$ \\
$\std$ & 1.7 \\
$\gamma$ (Discount Factor) & 0.99 \\
Initial Learning Rate & $2 \times 10^{-4}$ \\
Optimiser & Adam \\
$\lambda$ (GAE) & 0.95 \\
$\epsilon$ (PPO Clipping) & 0.3 \\
Maximum Gradient Norm & 0.5 \\
Entropy Coefficient & 0.05 \\
Value Coefficient & 1.0 \\
Number of Minibatches & 5 \\
Number of Update Epochs & 4 \\
Hidden Layer Dimensions & 50 \\
GRU Hidden Layer Dimension & 50 \\
Layer Norm & True \\
Activation Function & ReLU \\
Attention Embedding Dimension & 8 \\
\bottomrule
\end{tabular}
\end{table}

\begin{figure}[h]
    \centering
    \includegraphics[width=\textwidth]{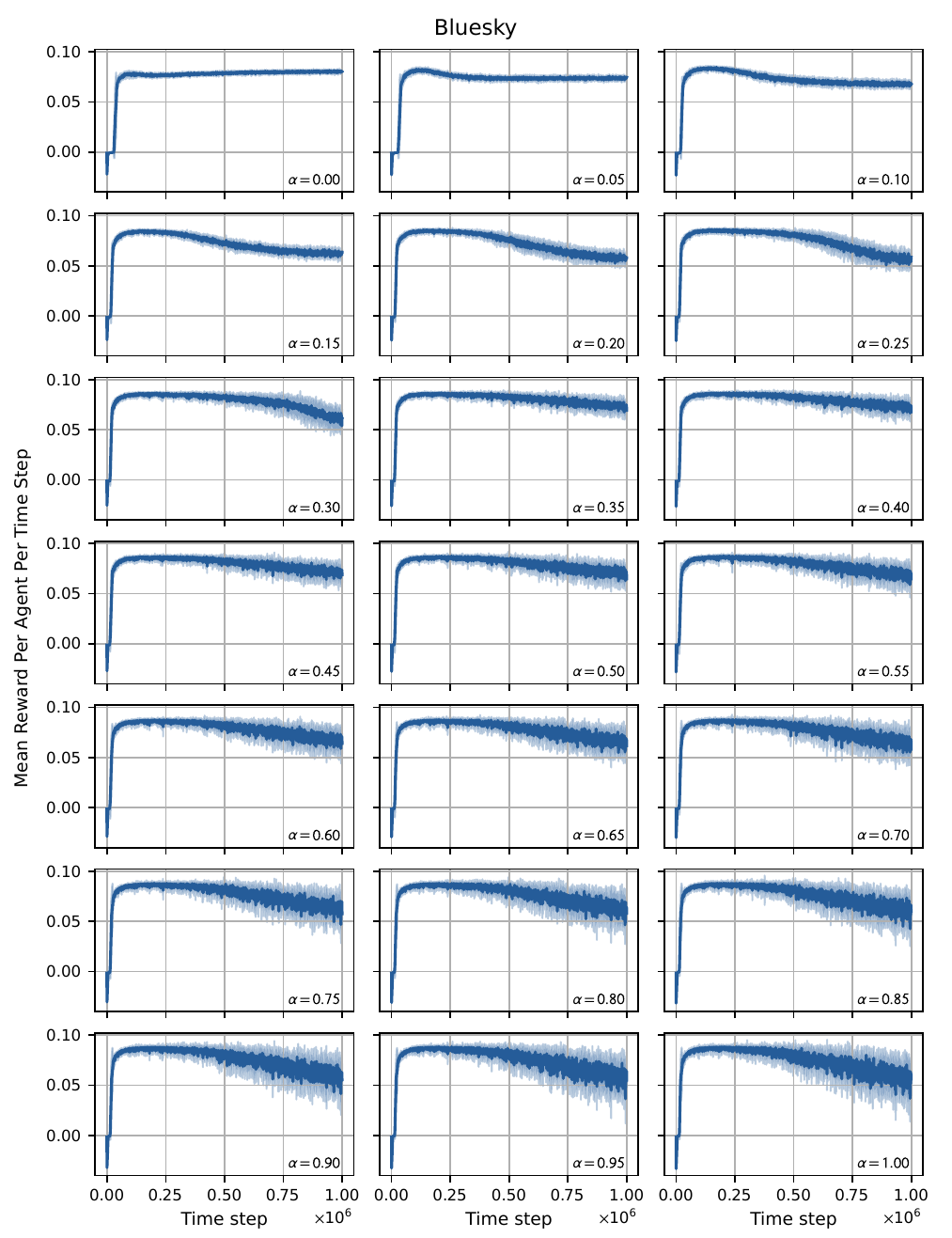}
    \caption{Reward training curves for the Bluesky network. Note the decrease in reward with time is due to the annealing of $p_\mathrm{signal}$.}
    \label{fig:training_curves_bluesky}
\end{figure}

\begin{figure}[h]
    \centering
    \includegraphics[width=\textwidth]{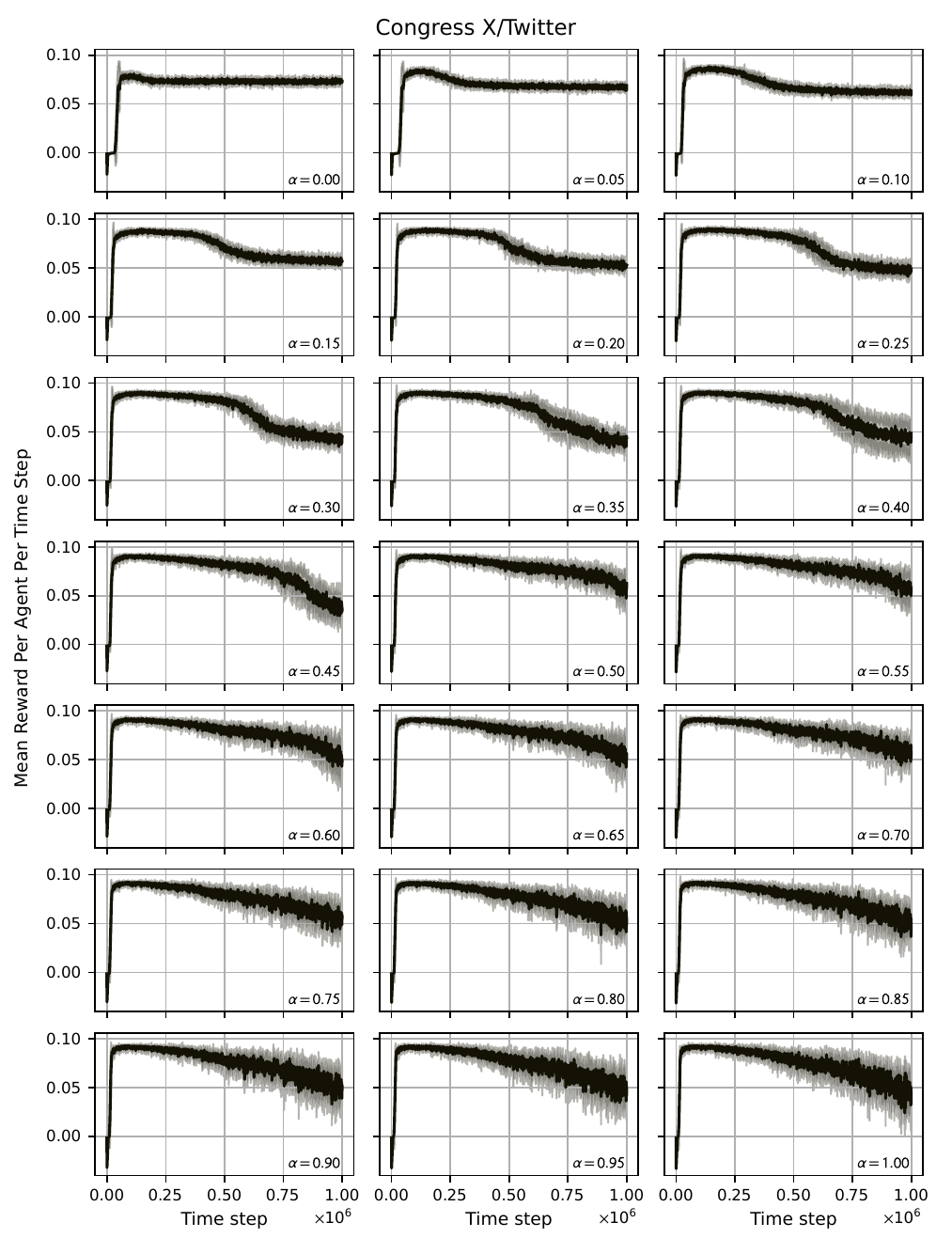}
    \caption{Reward training curves for the Congress network. Note the decrease in reward with time is due to the annealing of $p_\mathrm{signal}$.}
    \label{fig:training_curves_congress}
\end{figure}

\begin{figure}[h]
    \centering
    \includegraphics[width=\textwidth]{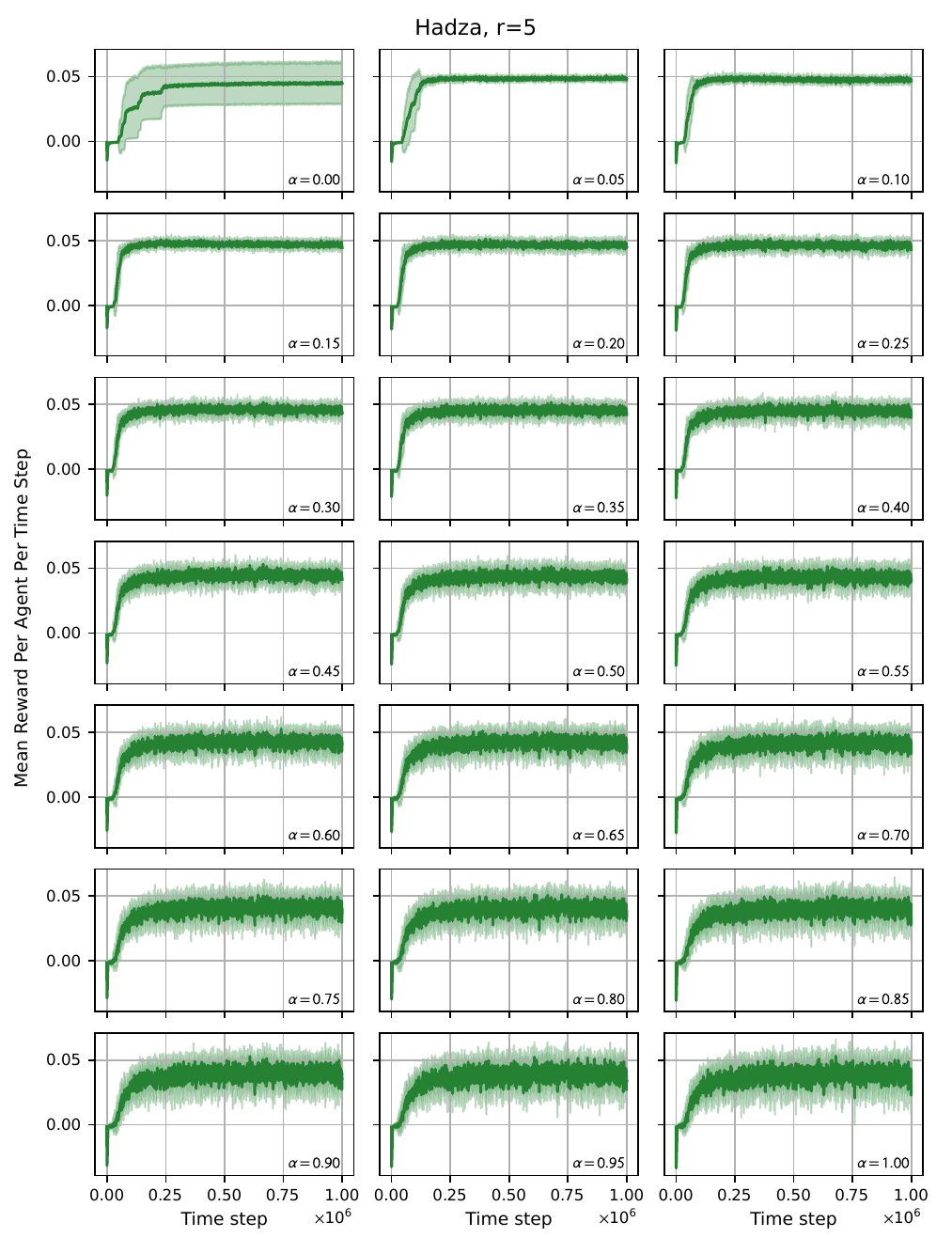}
    \caption{Reward training curves for the Hadza network with $r=5$.}
    \label{fig:training_curves_hadza_r5}
\end{figure}

\begin{figure}[h]
    \centering
    \includegraphics[width=\textwidth]{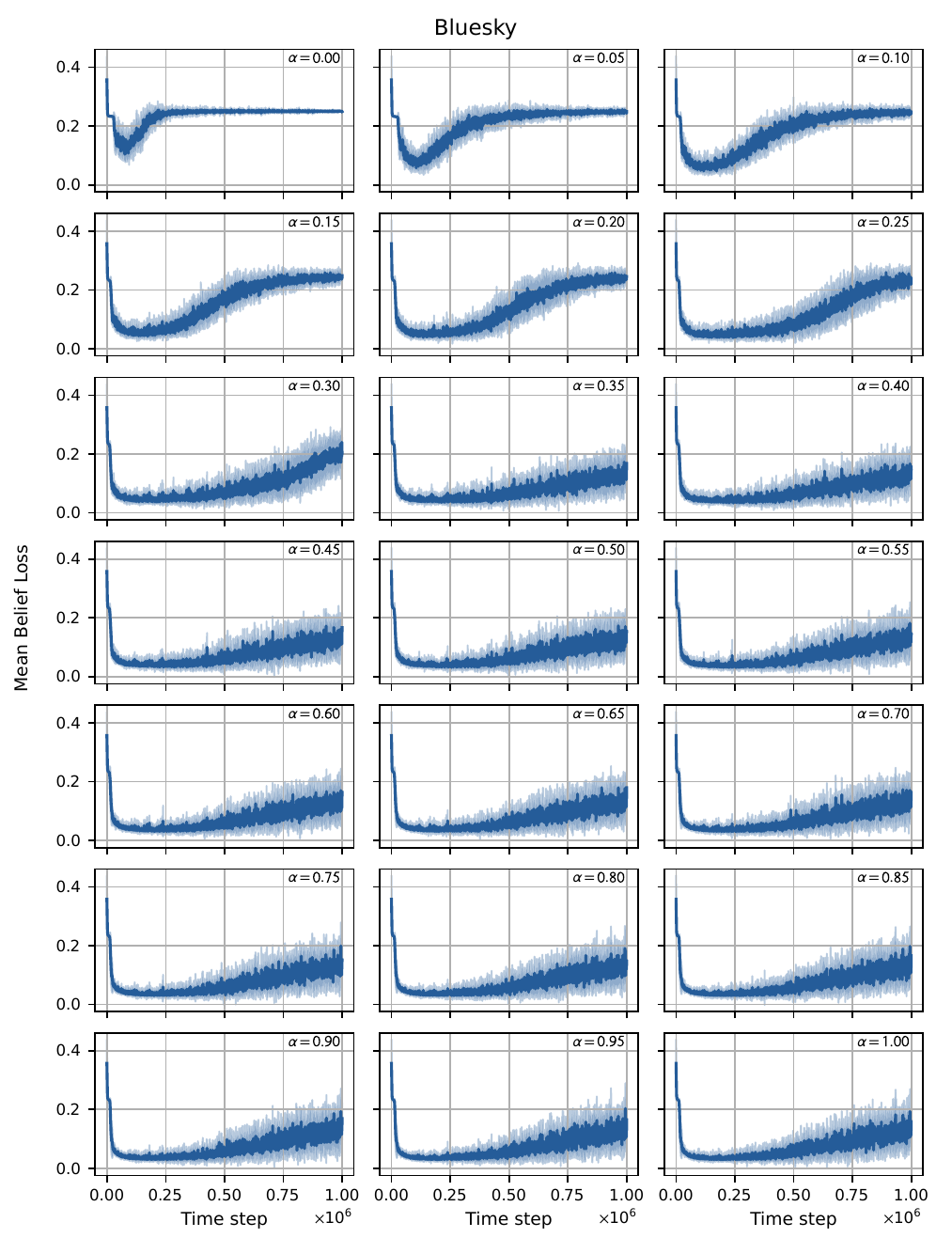}
    \caption{Belief loss training curves for the Bluesky network. Note the increase in loss with time is due to the annealing of $p_\mathrm{signal}$. Additionally, the large loss for small $\weight$ is due to a drop in population accuracy.}
    \label{fig:belief_loss_bluesky}
\end{figure}

\begin{figure}[h]
    \centering
    \includegraphics[width=\textwidth]{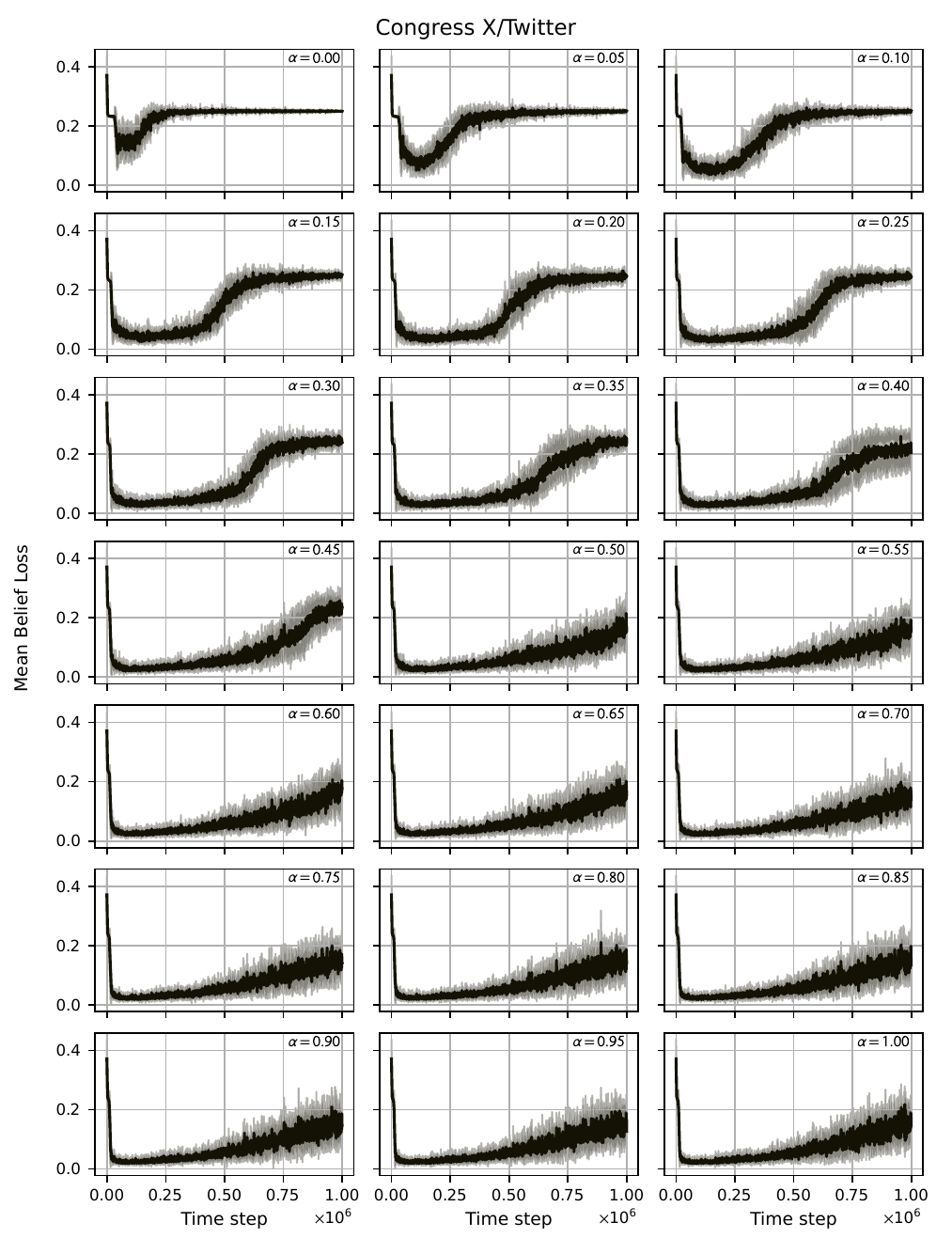}
    \caption{Belief loss training curves for the Congress network. Note the increase in loss with time is due to the annealing of $p_\mathrm{signal}$. Additionally, the large loss for small $\weight$ is due to a drop in population accuracy.}
    \label{fig:belief_loss_congress}
\end{figure}

\begin{figure}[h]
    \centering
    \includegraphics[width=\textwidth]{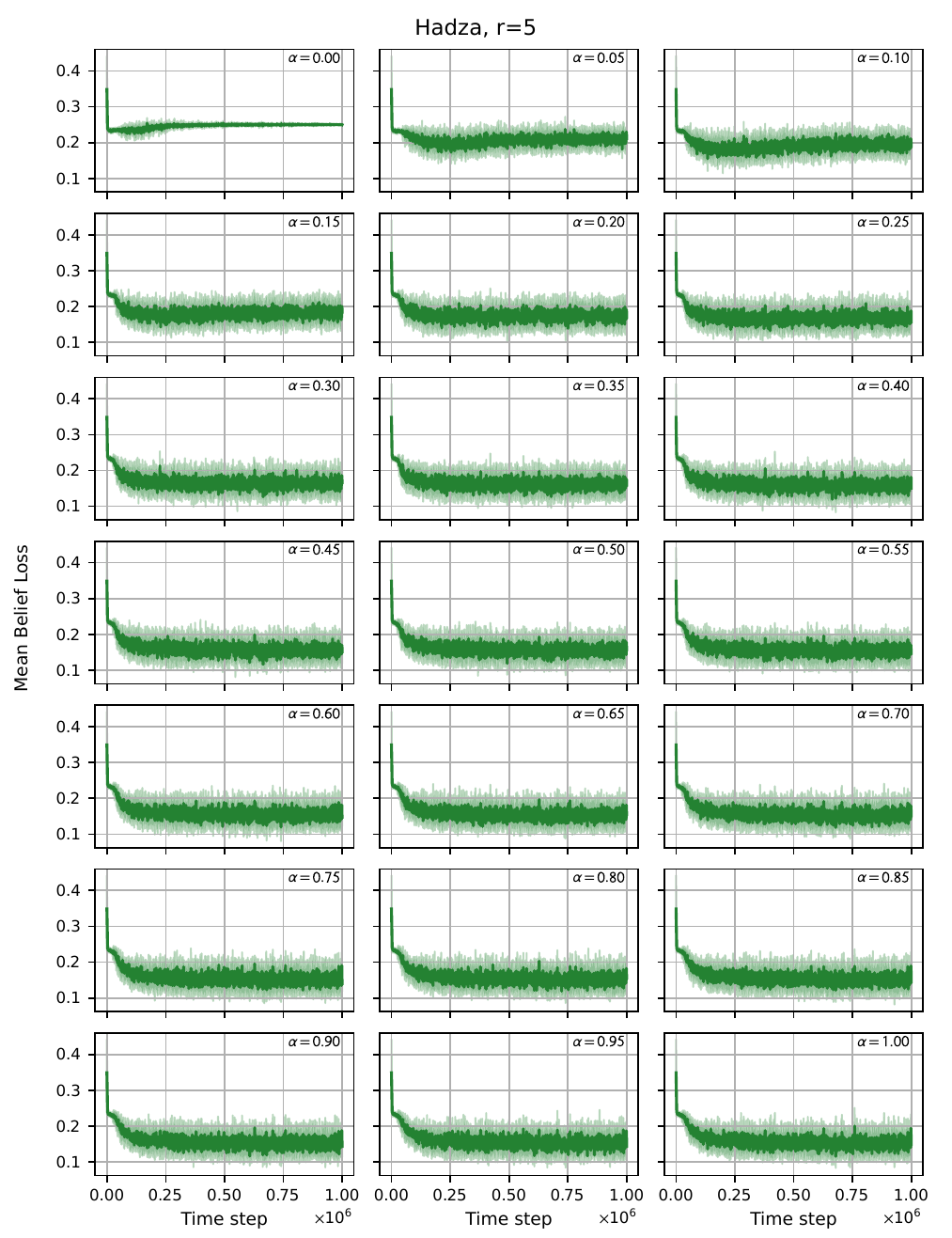}
    \caption{Belief loss training curves for the Hadza network with $r=5$.}
    \label{fig:belief_loss_hadza_r5}
\end{figure}

\FloatBarrier


\end{document}